\documentclass[oneside]{sn-jnl}


\usepackage{graphicx}%
\usepackage[strings]{underscore}
\usepackage{multirow}%
\usepackage{amsmath,amssymb,amsfonts}%
\usepackage{amsthm}%
\usepackage[title]{appendix}%
\usepackage{xcolor}%
\usepackage{textcomp}%
\usepackage{manyfoot}%
\usepackage{booktabs}%
\usepackage{algorithm}%
\usepackage{algorithmicx}%
\usepackage{algpseudocode}%
\usepackage{listings}%
\usepackage{lmodern}
\usepackage{url}
\usepackage[numbers]{natbib}

\begin{document}
        
\title[Article Title]{Flexible generation of daily Earth system model projections across radiative forcing scenarios}





\author*[1,2]{\fnm{Yu} \sur{Huang}}\email{y.huang@tum.de}

\author[1,2]{\fnm{Sebastian} \sur{Bathiany}}

\author[1,2]{\fnm{Shangshang} \sur{Yang}}

\author[1,2]{\fnm{Philipp} \sur{Hess}}

\author[1,2]{\fnm{Michael} \sur{Aich}}

\author*[1,2,3]{\fnm{Niklas} \sur{Boers}}\email{boers@pik-potsdam.de}

\affil[1]{\orgdiv{Munich Climate Center and Earth System Modelling Group}, \orgname{Department of Aerospace and Geodesy, School of Engineering and Design, Technical University of Munich}, \city{Munich}, \postcode{80333}, \country{Germany}}

\affil[2]{\orgdiv{Complexity Science}, \orgname{Potsdam Institute for Climate Impact Research}, \city{Potsdam}, \postcode{14473}, \country{Germany}}

\affil[3]{\orgdiv{Department of Mathematics and Statistics}, \orgname{University of Exeter}, \city{Exeter}, \postcode{EX4 4QF}, \country{United Kingdom}}

\abstract{Earth system model (ESM) projections of the climate system's response to anthropogenic forcing are central to assess the impacts of climate change and inform adaptation and mitigation policies. However, given their high computational cost, projections are only made for a limited set of standardized forcing scenarios with limited temporal extent, such as the Shared Socioeconomic Pathways (SSPs), the spatiotemporal resolution remains too low for direct impact assessments, and uncertainties cannot be comprehensively quantified. Recent data-driven models offer efficient and accurate high-resolution simulations for weather prediction, but cannot extrapolate to future greenhouse gas concentrations because they cannot capture the responses to unprecedented forcing, limiting their value for climate change projections.
Here, we combine response theory with a tailored generative machine learning framework to address this challenge. Our approach extracts the physical forced response to radiative forcing from monthly low-resolution ESM fields, and uses this response to guide a generative model to infer consistent daily global high-resolution temperature and precipitation projections. Our probabilistic approach generalizes across ESMs and provides long-term, bias-corrected responses to radiative forcing at high spatiotemporal resolution. It efficiently generates large ensembles needed for uncertainty quantification, effectively fills the gaps between existing SSPs, and readily extends climate projections to 2300 and beyond. Our framework hence complements ESM projections by providing efficient, stable, and high spatiotemporal resolution long-term climate projection ensembles across emission scenarios, enabling detailed impact assessment and exploration of long-term climate commitment.}

\maketitle
\newpage
\section*{Main}

In the context of ongoing anthropogenic climate change, high-resolution projections under changing greenhouse gas (GHG) concentrations are essential for assessing environmental and socioeconomic impacts of different emission scenarios, and for designing suitable adaptation and mitigation policies \cite{IPCC2022WGII_Full,  van2025exploring, palmer2019scientific}. High spatial and temporal resolution of modeled climate fields, such as for temperature and precipitation, is critical for driving impact models and capturing key processes and extremes \cite{eyring2023human, yoon2026extreme, ruane2015climate, craig2022overcoming}. At the same time, reliable climate risk assessment requires flexible generation of forcing scenarios \cite{van2025exploring, andrijevic2025representing, aengenheyster2018point, huisman2025projections, bevacqua2026moderate} and large ensembles of realizations for comprehensive uncertainty quantification \cite{bevacqua2026moderate, rodgers2021ubiquity, fischer2023storylines}.
Earth system models (ESMs) provide physical projections of climate variables under standardized forcing scenarios \cite{eyring2016overview, masson2021climate, IPCC_AR6_WGI_Chapter4_2021}, such as the shared socioeconomic pathways (SSPs), which describe plausible future socioeconomic developments and associated emissions \cite{masson2021climate, meinshausen2020shared, o2014new}. However, due to computational limitations, ESMs continue to have rather coarse resolution and projections are limited to small ensembles for small sets of emission scenarios. In addition, output fields from different ESMs usually differ in their model biases \cite{DoblasReyes2021, dinh2023revisiting} and their response to forcing \cite{knutti2017beyond,meehl2020context, rugenstein2020equilibrium}, making statistical bias adjustments against observational references necessary \cite{cannon2015bias, hess2022physically, qian2021projecting}. Similarly, to drive impact models with ESM projections, downscaling is required in addition to bias correction \cite{IPCC_AR6_WGI_Chapter4_2021, palmer2019scientific, palmer2019stochastic, hess2025fast,aich2025conditional}. Comprehensive coverage across possible forcing scenarios, while having large ensembles of accurate high-resolution ESM fields therefore remains a major challenge for climate change studies \cite{eyring2023human, palmer2019scientific, RN23}.\

A range of post-processing approaches have been developed to improve the representation of ESM climate fields. Conventional statistical methods, such as spatial interpolation and grid-point-wise quantile delta mapping (QDM) \cite{cannon2015bias, maraun2018statistical, gergel2024global}, remain widely used. For long-term climate simulations, these approaches aim to match the statistical properties of model outputs to observations. For example, QDM adjusts the probability distribution of ESM projections at each quantile, ensuring statistical consistency with observational references while preserving the climate change signal simulated by the ESM \cite{cannon2015bias}. More recently, generative machine learning (ML) approaches \cite{song2020score,creswell2018generative} have gained increasing attention for their ability to model complex high-dimensional distributions, improving both fine-scale spatial variability, multivariate dependencies, and the representation of extremes in post-processed ESM's climate fields \cite{hess2022physically,buster2024high,hess2025fast,lopez2025dynamical,aich2025conditional}. In this context, diffusion models \cite{song2020score, ho2020denoising} have emerged as a dominant paradigm for generative modeling, with strong performance, stability and flexibility, and the capacity to support diverse model variants \cite{song2020score, dhariwal2021diffusion, daras2024survey, chung2025diffusion, lipman2022flow}. Applying these approaches to weather forecasting has led to substantial gains in spatiotemporal variability and the efficiency of probabilistic ensemble prediction \cite{price2025probabilistic, yang2025generative,nai2025boosting}. The performance of diffusion models across downscaling \cite{hess2025fast, lopez2025dynamical}, data assimilation \cite{manshausen2025generative}, and weather prediction \cite{price2025probabilistic} indicated outstanding capability to learn the high-dimensional distribution of climate fields.  \

Despite these advances, existing approaches remain constrained by their reliance on precomputed ESM outputs. Statistical interpolation to new radiative forcing scenarios is inherently limited, as such approaches cannot preserve plausible spatiotemporal variability. Likewise, generative ML-based downscaling cannot generate scenarios for which no ESM simulation data exists \cite{hess2025fast, aich2025conditional, lopez2025dynamical}, due to the lack of explicitly simulated responses to radiative forcing. 
The above calls for a combination of physical estimates of the response to forcing and generative ML for efficient, realistic, and high-resolution spatiotemporal variability. Indeed, in recent years, ML–based parameterizations and hybrid modeling approaches have partially mitigated computational costs by substituting computationally expensive components of ESMs with efficient ML operators \cite{kochkov2024neural, ma2026mloz, gelbrecht2023differentiable, palmer2019stochastic}. However, sufficiently comprehensive hybrid ESMs that at least partially address the above challenges are still lacking \cite{moses2025dj4earth, kochkov2024neural,watt2025ace2}. \

Ruelle’s response theory \cite{ruelle1998general, ruelle2009review} provides an alternative, physically grounded method for modeling the response of the climate system to external forcing \cite{ragone2016new, lucarini2017predicting, lucarini2023theoretical}. By introducing Green’s functions to represent the system’s response across relevant time scales, the approach defines a causal operator that maps external forcing to the resulting climate response \cite{lucarini2017predicting, lucarini2023theoretical}. Green’s functions estimated from idealized simulations of a given ESM have been shown to retain the model’s intrinsic sensitivity to forcing \cite{ragone2016new, aengenheyster2018point, lucarini2017predicting, lucarini2024detecting}. Consequently, the approach can effectively reproduce the gradual responses and long-term trends of climate variables across a range of forcing scenarios \cite{ragone2016new, lucarini2017predicting, lembo2020beyond, womack2025rapid, sandstad2025meteorv1}, although it does not resolve fine-scale spatiotemporal variability, remaining limited to smooth spatial fields at annually or monthly temporal resolution. \

In this study, we bridge the gap between efficient machine-learning generation of weather fields, and flexible estimation of the physical response to broadly varying radiative forcing scenarios. As illustrated in Fig. 1 and detailed in the Methods, we introduce a response theory–informed generator (RIG) approach that combines the strengths of generative machine learning with Ruelle’s response theory. Guided by response patterns derived from monthly coarse-resolution fields of a given ESM, RIG produces global daily fields of precipitation and temperature at 0.25° spatial resolution across radiative forcing scenarios. This approach preserves the ESM's responses to forcing while capturing fine-scale spatiotemporal variability and statistical characteristics consistent with observational reference data. In particular, our approach provides a principled solution for generating high-resolution daily climate fields for ad hoc defined forcing scenarios that have not been provided by process-based ESMs. \

\subsection*{Response pattern}
We first assess the transferability of response theory (RT) across anthropogenic forcing scenarios using temperature and precipitation fields from the state-of-the-art Earth system model MPI-ESM1.2-LR. Grid-point-wise Green’s functions are estimated from monthly mean climate variables from the abrupt CO$_2$ quadrupling (4×CO$_2$) experiment (Methods and Supplementary Note 2). When driven by effective radiative forcing from previously unseen scenarios, the resulting RT model accurately reproduces monthly temperature and precipitation responses well, as illustrated for the 1\% yr$^{-1}$ CO$_2$ increase (1pctCO$_2$) experiment (Supplementary Fig. S1).\

We next extend RT to SSP scenarios, which involve multiple forcing components with distinct radiative contributions \cite{dentener2021annex, Smith2021IPCC_AR6_AnnexIII} (Supplementary Fig. S2 and Table S1). Applying RT with greenhouse gas (GHG) forcing alone leads to systematic overestimation of both temperature and precipitation responses, reflecting the missing cooling effect of aerosols (Supplementary Figs.S5 and S6). To account for this, we explicitly estimate Green’s functions for anthropogenic aerosol forcing using effective radiative forcings and response signals from SSP2-4.5 (see Supplementary Note 3 for details). Incorporating both GHG and aerosol responses yields a comprehensive RT model that captures the combined radiative forcing effects and reproduces response patterns across SSP scenarios not used in its construction (Supplementary Figs. S3 and S4). We note that SSP scenarios also include additional forcing components associated with land-use changes and volcanic eruptions \cite{Smith2021IPCC_AR6_AnnexIII}. We did not explicitly estimate the corresponding Green’s functions for these components. Volcanic forcing is incorporated directly into the aforementioned response theory model (Supplementary Note 3). The global impacts of localized land-use changes are relatively minor compared to other forcing components and are not considered here. The feasibility of RT is not limited to a single ESM. Our examination using two additional state-of-art ESMs, CESM2 and IPSL-CM6A-LR, demonstrates that the RT model can be constructed in a consistent manner and effectively reproduces response patterns across all available SSP scenarios (Supplemenatry Figs. S5 and S6). This finding on the RT generalizability across ESMs is consistent with recent studies that were based on annual mean data \cite{womack2025rapid, sandstad2025meteorv1}. \

In this context, we note that RT may offer a practical means to generate responses to any other potential radiative forcing scenario and, for example, complement missing SSP scenarios within the Coupled Model Intercomparison Project Phase 6 (CMIP6) \cite{o2016scenario, tebaldi2021climate, eyring2016overview}. For example, simulations for SSP4 are not available for MPI-ESM1.2-LR (Supplementary Fig. S5a). The RT-predicted responses exhibit consistent temporal trends with the MPI-ESM1.2-LR simulations across both the historical period (1850–2014) and the future period (2015–2100) under SSP1, SSP2, SSP3, and SSP5 (Supplementary Figs. S5 and S6). Furthermore, when driven by effective radiative forcings corresponding to the SSP4 scenarios (Supplementary Fig. S2), RT can deliver temperature and precipitation responses for SSP4-3.4 and SSP4-6.0 (Fig. 2), thereby extending the scenario coverage beyond that available in the original MPI-ESM1.2-LR simulations. \

Despite these strengths, the temperature and precipitation fields predicted by RT (e.g., Fig. 3a) remain limited by the monthly temporal resolution of the climate variables and the native spatial resolution of the underlying ESMs, which is approximately 1.875° in the case of MPI-ESM1.2-LR. This limitation motivates the need for high-resolution generation approaches in both space and time. \

\subsection*{Response pattern guided generation} 
Building on the forced climate response, we aim to incorporate daily weather variability into the generated fields. We use the response theory outputs to guide a generative diffusion model \cite{song2020score} that we trained beforehand to produces consecutive, time- and variable-consistent daily 0.25° fields of temperature and precipitation jointly \cite{huang2026generative} (Fig. 1). The diffusion model initializes from stochastic noise and, through multi-step iterative denoising with a neural network, generates daily climate fields ($\mathbf{X}_{d}$) jointly with the corresponding monthly fields ($\mathbf{X}_{m}$). At each denoising step, we also provide the daily fields from the preceding two days, $\mathbf{X}_{d-2}$ and $\mathbf{X}_{d-1}$ as prior input to the neural network (Methods). To train this model, one can utilize simulation data from a specific ESM or, alternatively, observational reference data. Here, to enable compatibility with multiple ESMs and meet the demand for high spatial resolution on a globally regular grid, our training (Methods) is conducted using the ERA5 daily reanalysis data (1979–2014), which has been assimilated with real-world observations and is commonly used as reference for tuning and evaluation of ESMs \cite{hersbach2020era5, di2025evaluation, hess2022physically}.
At the inference stage, following conditional guidance techniques in a Bayesian formulation \cite{song2020score, chung2022diffusion}, the known monthly fields act as conditioning information (Methods and Supplementary Notes 6-7) and guide the trained diffusion model to generate consecutive daily 0.25° fields.  \ 

We present RIG-generated fields based on the forced responses of MPI-ESM1.2-LR for the year 2080 under the SSP3-7.0 scenario. The monthly fields (Fig. 3a) derived from response theory are used to guide the generation of consecutive daily fields at 0.25° resolution (Fig. 3b). Fig. 3d shows a comparison of the RIG-generated time series with the native daily time series from MPI-ESM1.2-LR, shown as global averages. Meanwhile, RIG generates probabilistic ensembles, as stochastic sampling in the diffusion process produces unique realizations drawn from the learned distribution \cite{song2020score, price2025probabilistic, nai2025boosting}. The 100 RIG ensemble members exhibit mean-state levels and seasonal variations consistent with MPI-ESM1.2-LR daily time series (Fig. 3d). This result reflects two capabilities of RIG: the response theory model captures the seasonal cycle and mean-state responses, while the guided diffusion model inherits these signals and their associated mean-state trends during daily field generation (Supplementary Note 7) \cite{hess2025fast, aich2025conditional, huang2026generative}. Furthermore, the RIG ensemble spread for temperature largely spans that of MPI-ESM1.2-LR, although certain modes of large-scale variability remain insufficiently represented. This reflects the fact that neither response theory nor the diffusion model is explicitly designed to capture ESM-specific internal variability during training. The RIG ensemble spread in precipitation encompasses the ESM outputs, suggesting that precipitation variance is contributed primarily from weather-scale fluctuations, which are captured by the diffusion model. \

The applications of RIG to CESM2 and IPSL-CA6M-LR yield similar performances (Supplementary Figs. S9-S10). Although three ESMs show different climate mean states and seasonality, RIG-generated daily data reproduces this diversity, demonstrating that RIG is adaptive to the different ESMs. \

\subsection*{Generating daily climate fields with observational characteristics} 
Our goal for RIG is to directly generate high-resolution daily climate fields that have as little bias as possible, i.e., are statistically consistent with observational characteristics. To this end, we first apply QDM to calibrate the distributions of temperature and precipitation fields predicted by the RT model, aligning them with ERA5 reanalysis over the 1979–2014 period while preserving the long-term climate change signal. RIG then uses the QDM-calibrated monthly coarse-resolution response patterns to guide the generation of daily fields at 0.25° horizontal resolution. For applications based on MPI-ESM1.2-LR, we refer to this configuration as RIG-MPI. The period 2015–2023 is withheld entirely from both QDM calibration and diffusion model training, and is used as independent test data to evaluate the statistical properties of RIG-MPI outputs. As the effective radiative forcing and mean climate states across different SSP scenarios have not yet diverged during this period (Fig. 2 and Supplementary Fig. S2), we present evaluation results for the 2015–2023 period under SSP3-7.0 as a representative example. \ 

The RIG-MPI outputs reproduce the latitudinal profiles and probability distributions of daily temperature and precipitation in close agreement with the observational reference (Fig. 4). In comparison, the MPI-ESM1.2-LR's native daily temperature also shows good agreement with the observational reference in these two statistical aspects (Figs. 4a and 4b). However, its daily precipitation exhibits discrepancies relative to the observational reference in both latitudinal profile and probability distribution (Figs. 4d and 4e).
After applying conventional spatial interpolation and QDM postprocessing to the MPI-ESM1.2-LR outputs, the latitudinal profile of precipitation is improved (Fig. 4d). However, the raw ESM output systematically underestimates the frequency of heavy precipitation events exceeding $40\,\mathrm{mm\,day^{-1}}$, and this bias persists after standard QDM correction (Fig. 4e). This suggests that the integration of a diffusion model in RIG-MPI improves the representation of daily precipitation extremes compared to ESM itself. \

The daily fields of RIG-MPI also exhibit spatiotemporal variability that is close to the observational reference. Spatial power spectrum analysis shows that the daily temperatures simulated by MPI-ESM1.2-LR are consistent with the observational reference in terms of spatial variability (Fig. 4c), whereas the spatial variability of precipitation is underestimated (Fig. 4f). In addition, the power spectra do not capture wavelengths below 400 km, owing to the limited spatial resolution of the original ESM data.
Spatial interpolation and QDM postprocessing can help compensate for the lack of short-wavelength spatial structure in daily temperature fields (Fig. 4c), but they provide limited improvement for the spatial variability of precipitation (Fig. 4f). This is because QDM is applied independently at each grid cell and for each variable, and does not account for spatial structure or inter-variable dependencies. RIG-MPI, meanwhile, produces temperature and precipitation fields that more closely align with the observational reference across all resolved wavelengths (Figs. 4c and 4f). \ 

Analysis of autocorrelation in deseasonalized daily time series further reveals that raw simulations from MPI-ESM1.2-LR overestimate temporal persistence for both temperature and precipitation over tropical regions (Fig. 4h), with similar behavior also found in simulations from CESM2 and IPSL-CM6A-LR (Supplementary Fig. S11). By comparison, the spatial patterns of autocorrelation in RIG-MPI outputs closely resemble those of the ERA5 reference (Fig. 4j). In the cross-correlation analysis between temperature and precipitation, we further find that, compared with the original ESMs, the RIG-generated daily climate fields exhibit spatial distributions of cross-correlations that are more consistent with those in the ERA5 reference (Supplementary Fig. 12). These results indicate that RIG can complement ESMs in generating daily climate fields with higher spatial resolution and statistical properties closer to observational references. \

\subsection*{Generating daily climate fields under varying radiative forcing} 

Guided by monthly fields derived from response theory, RIG generates consecutive, temporally consistent daily fields from 2015 to 2100 for a given scenario and is able to reproduce long-term climate trends under changing radiative forcing. Under SSP3-7.0, for example, linear temperature trends at the grid-cell level over 2050–2100 (Fig. 5b) reveal the global warming signal, with amplified warming in the Arctic relative to lower latitudes. RIG-MPI also captures the distribution of areas with increasing and decreasing tropical precipitation. These spatial patterns match those simulated by MPI-ESM1.2-LR (Fig. 5a), while global means increase consistently over 2015–2100 (Fig. 5d). This consistency arises from response patterns derived from MPI-ESM1.2-LR, with the guided diffusion model preserving the underlying forced signals during generation. \

A benefit of our method is that it can readily generate temperature and precipitation projections in response to radiative forcing scenario coverage beyond the scenarios available in the CMIP6 archive, such as SSP4-3.4 and SSP4-6.0 for MPI-ESM1.2-LR. Under SSP4-3.4, the anthropogenic climate change signal is markedly weaker than under SSP3-7.0, with pronounced warming largely confined to polar regions (Fig. 5c). Annual mean precipitation changes are correspondingly modest and remain localized, primarily east of the Pacific warm pool. Fig. 5d summarizes the long-term trends of global mean temperature and precipitation across seven SSP scenarios generated by RIG-MPI. Because response patterns have been bias-corrected before the generation, RIG-MPI outputs exhibit different mean states relative to MPI-ESM1.2-LR (Fig. 5d), most evident in global mean precipitation. We also perform our assessment based on other ESMs from CESM2 and IPSL-CM6A-LR, namely RIG-CESM2 and RIG-IPSL, with similar results (Supplementary Figs. S13). This capability allows RIG to generalize across radiative forcing scenarios, without the need to rerun the ESMs. Moreover, it is straightforward to employ RIG to extend existing projections beyond the typical end year in 2100, e.g. using the extended SSP emission scenarios \cite{meinshausen2020shared, dentener2021annex,masson2021climate}; we demonstrate this by extending the standard SSP1-2.6, SSP2-4.5 and SSP5-8.5 projections from 2100 until 2300 for the IPSL-CM6A-LR (Supplementary Note 8 and Fig. S17). For the extended SSP1-2.6 scenario, native IPSL-CM6A-LR simulations beyond 2100 are available from CMIP6, and the RIG-generated extension exhibits a consistent long-term trend with these native simulations. \

In addition, we investigate the changes in the standard deviation and extreme values of RIG-MPI generated daily data at each grid cell over 2065–2100 relative to 1979–2014. This is motivated by the fact that in the real world, it is not only the climate mean that changes due to forcing, but also the variability around the mean \citep{easterling2000climate, bathiany2018climate, sippel2020climate}. Under SSP3-7.0, RIG-MPI and MPI-ESM1.2-LR exhibit consistent changes in temperature standard deviation across several regions (Supplementary Figs. S14a and S14c): The standard deviation of temperature decreases in the Arctic, along the Antarctic coastal regions, and over South Asia, while it increases over North America, the North Atlantic, and the Amazon. RIG-generated daily fields also exhibit changes in hot extremes, with patterns similar to MPI-ESM1.2-LR (Supplementary Figs. S15). We emphasize that the diffusion model is not trained on the internal variability of ESMs under future radiative forcing scenarios. We hence conclude that the ability of RIG to produce forced changes in climate variability arises from learning the observed relationship between monthly mean states and daily variability into the future.  \

\section*{Discussion}
We introduced a framework that integrates response theory with generative machine learning to enable efficient and accurate generation of high spatio-temporal resolution climate fields. RIG extends ESM-based climate projections in four aspects. First, it enables the generation of bias-corrected, high-resolution daily fields under flexible radiative forcing scenarios beyond standard SSP scenarios without requiring additional costly ESM simulations or retraining, thus effectively filling the projection space between the small set of standard emission scenarios. Second, RIG enables efficient ensemble boosting, readily generating large projection ensembles for a given forcing scenario to enable thorough uncertainty quantification. Third, the generated fields are inherently bias corrected and downscaled in time and space, providing daily global fields at 0.25° horizontal resolution with accurate spatiotemporal variability and representation of extremes, which can be readily employed for impact assessments. Forth, our framework can be used to extend existing projections to study long-term climate change commitments, which we exemplify by extending standard SSP runs until 2300.   \

In particular, RIG provides an avenue for analyzing the extreme events and quantify their uncertainty. As an illustrative example, we examine the day with the most intense precipitation over a region in eastern Asia (Supplementary Table S2) during 2065–2100 under the SSP3-7.0 scenario, comparing spatial precipitation fields from MPI-ESM1.2-LR (Fig. 6a), and their RIG counterparts (Fig. 6b). 
RIG-MPI reduces the blurriness of native ESM precipitation fields, producing sharper and more spatially coherent structures. 
Moreover, two realizations of RIG-MPI exhibit diverse spatiotemporal evolutionary patterns during their respective days with the most intense precipitation (Fig. 6b). This ensemble spread captures the intrinsic stochasticity of extreme precipitation, which in turn shapes the uncertainty of return-period distributions (Fig. 6c), which increases with event intensity. Fig. 6c further compares return-period distributions of regional precipitation events derived from native ESMs and RIG realizations. Over the historical period 1979–2014, MPI-ESM1.2-LR and IPSL-CM6A-LR overestimate return periods for moderate to extreme precipitation relative to the ERA5 reference, while CESM2 underestimates them. In contrast, RIG realizations for all three ESMs produce return-period distributions consistent with ERA5. In the future period 2065–2100, the ensemble spread of RIG-MPI shifts distinctly relative to the historical level, indicating shortened return periods for intense precipitation events under the SSP3-7.0 scenario. These results underscore the potential of generative machine learning as a tool for generating large ensembles of extreme weather events under specific climatological conditions \cite{price2025probabilistic, nai2025boosting, antonio2026seasonal}. The ability to generate large ensembles of high-resolution climate projections is crucial for subsequent uncertainty-aware hydrological, ecological, or socioeconomic impact assessments. \

Several avenues for future development could further enhance RIG's utility for anthropogenic climate change projection and impact assessment. For example, incorporating more climate variables into RIG’s training may enable the generative ML model to provide more comprehensive climate realizations, better capturing compound extreme events and multiscale spatiotemporal interactions. In addition, because traditional Quantile Delta Mapping applies bias correction independently to each climate variable at individual grid cells, generative ML–based approaches \cite{hess2022physically} could be explored as an alternative to better capture multivariate joint distributions in the bias correction, and to improve the spatial coherence of bias-corrected fields \cite{dinh2023revisiting, cannon2018multivariate, cannon2015bias}. Moreover, RIG currently produces climate fields at 0.25° spatial resolution, but scaling to higher resolutions, such as 0.1°, is technically feasible and would support additional applications while aligning with emerging observational and modeling products \cite{brenowitz2025climate}.In particular, integrating multi-source observational precipitation data \cite{sun2018review} is essential for improving the model’s ability to realistically represent precipitation \cite{sun2025fusion}. Meanwhile, generating one year of global daily temperature and precipitation fields at 0.25° resolution with our current generative model requires approximately 7,665 seconds; recent advances in generative ML, particularly for diffusion models, have enabled faster sampling and improved guided generation \cite{lipman2022flow, hess2025fast, chen2024diffusion, li2025back}. These techniques hold promise for integration into the RIG framework in the future, potentially enhancing both performance and efficiency.

Another consideration is the further advancement of response theory. In this study, we apply a first-order Green's function approximation (Methods). This linear treatment may underrepresent nonlinear feedbacks in the climate system \cite{lucarini2017predicting}, and assumes that external forcing keeps the system sufficiently far from tipping points \cite{lembo2020beyond, lucarini2024detecting}, as the linear Green's function can diverge when strong positive nonlinear feedbacks are included. Incorporating higher-order Green’s functions in future work may further improve the model’s ability to represent nonlinear climate responses \cite{lucarini2023theoretical}. Another limitation is that we estimate the aerosol-specific Green’s function using the residual component from the SSP2-4.5 scenario after removing the GHG-driven response (Supplementary Note 3). Moreover, we do not infer aerosol-driven response characteristics from dedicated aerosol-forcing ESM experiments, nor do we account for the potentially heterogeneous spatial distribution of aerosols, which may lead to uncertainties in the estimated local aerosol responses. Our approach also does not explicitly consider effects of land-use change, although their equivalent in terms of radiative forcing is included. These limitations should be further evaluated and refined using more targeted ESM experiments in future work. Furthermore, it is of interest to explore implementations of response theory directly within generative modeling, as recent work has identified connections with the generalized fluctuation–dissipation theorem ~\cite{giorgini2024response, giorgini2025predicting}.

We introduced a combination of response theory and generative diffusion models to efficiently generate large ensembles of accurate high-resolution daily climate projections from coarse-resolution monthly ESM fields across radiative forcing scenarios. These are essential for better addressing the socioeconomic and ecological challenges of anthropogenic global warming in an uncertainty-aware manner. In recent years, generative machine learning has achieved notable success in weather forecasting and is increasingly advancing bias correction and downscaling of climate modeling projections. Enabling generative modeling to produce physically plausible responses to changing external forcing offers an efficient avenue for integrating insights from forced ESM responses with high-resolution observational constraints. Our physical climate response guided generative modeling approach can be employed for a broad range of applications, including paleoclimate reconstructions \cite{haywood2016pliocene}, inverse problems for anthropogenic climate change \cite{hegerl1996detecting, terhaar2022adaptive, eyring2023human}, as well as studies directly relevant for IPCC assessments, such as climate change attribution \cite{fischer2021increasing, hegerl1996detecting}, climate change impacts assessment \cite{buster2024high, ruane2015climate}, and investigation of long-term climate change committments via extensions of SSP-based climate projections\cite{masson2021climate,beyond2100}.

\section*{Methods}
\subsection*{Data}
\subsubsection*{\normalfont Climate fields data}
We use surface air temperature and total precipitation data from the MPI-ESM1.2-LR simulations archived in the CMIP6, including the abrupt \text{4$\times$CO$_2$}, \text{1pctCO$_2$}, historical (1850--2014), pre-industrial and SSP-scenario (2015--2100) experiments \cite{eyring2016overview, tebaldi2021climate}. The MPI-ESM1.2-LR data \cite{mauritsen2019developments} has a daily temporal resolution and a spatial resolution of approximately $1.875^\circ \times 1.875^\circ$, and are aggregated to monthly means for the subsequent estimation of the response theory operator. For comparison, corresponding simulation data from CESM2 \cite{danabasoglu2020community}, with a horizontal resolution of $1.25^\circ \times 0.94^\circ$, and from IPSL-CM6A-LR\cite{boucher2020presentation}, with a horizontal resolution of $1.25^\circ \times 2.5^\circ$, are processed in the same manner.

ERA5 is the global atmospheric reanalysis dataset produced by the European Centre for Medium-Range Weather Forecasts (ECMWF) through the assimilation of multi-source Earth observations \cite{hersbach2020era5}, providing spatiotemporally continuous climate variables on a regular grid. Here we use daily surface air temperature and total precipitation from the ERA5 dataset as training target for the diffusion model. The ERA5 data has an original hourly temporal resolution and a horizontal resolution of $0.25^\circ \times 0.25^\circ$, covering the period 1979–2023. For this study, the ERA5 fields are aggregated to daily values. 

\subsubsection*{\normalfont Effective radiative forcing}
In the \text{abrupt 4$\times$CO$_2$} experiment, the atmospheric CO$_2$ concentration is instantaneously quadrupled, whereas in the \text{1pctCO$_2$} experiment, \text{CO$_2$} concentration increases by 1\% per year until it reaches four times the pre-industrial level. For these \text{CO$_2$} experiments, the effective radiative forcing is a logarithmic function of atmospheric \text{CO$_2$} concentration \cite{meinshausen2020shared, masson2021climate} (Supplementary Note 1). For both historical and future scenarios under various SSP pathways, the effective radiative forcings for all greenhouse gas and aerosol components are provided by Intergovernmental Panel on Climate Change (IPCC) Working Group I \cite{dentener2021annex, Smith2021IPCC_AR6_AnnexIII} (Supplementary Table S1).

\subsection*{Response theory model} 
Ruelle's response theory \cite{ruelle1998general, ruelle2009review} describes the expectation value of a forced system variable as a perturbative expansion:
\begin{equation}
\ x(t)
= x_{0} + \sum_{i=1}^{\infty} \Delta x^{(i)}(t).
\end{equation}

Here $x_{0}$ denotes the expectation value of a climate variable in the pre-industrial period. The anomalous change $\Delta x^{(i)}(t)$ at time $t$ relative to the pre-industrial level, are the result from convolution integrals of the anomalous effective radiative forcing $\Delta f(t)$ relative to the pre-industrial effective radiative forcing. Previous studies \cite{ragone2016new, lucarini2017predicting, lembo2020beyond, lucarini2023theoretical, lucarini2024detecting, sandstad2025meteorv1} show that the first-order item can explain the climate variables' response to the GHG forcing in ESMs, that is resolved as: 
\begin{equation}
\ \Delta x^{(1)}(t)
=\int_{-\infty}^{\infty}G^{(1)}(\tau) \Delta f(t-\tau) d\tau.
\end{equation}
$G^{(1)}(\tau)$ is the first-order Green's function of the climate variable, which accounts for the system response effects at individual time scales $\tau$. Here, we estimate the Green’s functions for GHG and aerosols separately.

In the \text{abrupt 4$\times$CO$_2$} experiment, the temporal variation of effective radiative forcing follows a Heaviside function, with a constant value of anomalous effective radiative forcing $\Delta f_{4\times CO_2}$. This enables to differentiate Eq. (2) and gives \cite{ragone2016new, lucarini2023theoretical}: 
\begin{equation}
\  G^{(1)}(\tau)
= \frac{1}{\Delta f_{4\times CO_2}} \frac{d}{d\tau} \Delta x.
\end{equation}
The Green’s function is thus obtained by approximating temporal derivatives of the climate variable time series. We estimate GHG-specific Green’s functions for temperature and precipitation separately at each grid cell using data from the \text{abrupt 4$\times$CO$_2$} experiment. For each calendar month (January to December), we construct temperature or precipitation anomalies relative to the corresponding pre-industrial level using data from that calendar month across years, and compute the Green’s functions accordingly (See Supplementary Note 2 for details). For other forcing scenarios, the GHG-driven response of the climate variables at each grid cell can be predicted by setting the derived GHG-specific Green’s functions together with the time series of GHG effective radiative forcing  into Eq. (2) (Supplementary Note 2).

For the SSP scenarios, effective radiative forcings from individual GHG components are aggregated into a sum forcing time series, which is then integrated with the GHG-specific Green's function to predict temperature and precipitation responses associated with GHG forcing.
Special consideration is given to effective radiative forcings associated with changes in anthropogenic aerosol concentrations. We use the ESM simulation data from SSP2-4.5 scenario to estimate the aerosol-specific Green's function (Supplementary Note 3). As a result, the response terms associated with GHGs and aerosols are incorporated into the response theory framework, and subsequently used to predict climate responses under other unseen SSP scenarios.

\subsection*{Generative modeling for climate fields}

To formulate the generative learning problem for climate fields, we first define the state variables and their probabilistic structure as follows. The daily global longitude--latitude maps of climate variables (temperature and precipitation from ERA5 in this study) are represented by $\mathbf{X}_d$, while their corresponding 30-day averaged fields centered on day $d$ are denoted by $\mathbf{X}_m$. 
The combined state is defined as $\mathbf{X} := (\mathbf{X}_d, \mathbf{X}_m)$, where $\mathbf{X} \in \mathbb{R}^{n}$ and $n$ is the data dimension. The initial state $\mathbf{Z}$ is usually sampled from Gaussian distribution \text{$\mathcal{N}(0, I)$}, where $\mathbf{Z}, I \in \mathbb{R}^{n}$. In addition, the daily fields from one and two days prior to the day \textit{d} are denoted by $\mathbf{X}_{d-1}$ and $\mathbf{X}_{d-2}$, respectively. The generative model is formulated as an estimation of the following conditional probability distribution: 
\begin{equation}
    P(\mathbf{X}_{d}, \mathbf{X}_{m} \mid \mathbf{Z}, \, \mathbf{X}_{d-1}, \mathbf{X}_{d-2}), \, \, \text{$\mathbf{Z} \sim \mathcal{N}(0, I)$}.
\end{equation}
As a result, initialized from Gaussian noise samples, generative ML jointly infers the daily and monthly fields conditioned on the two preceding daily fields. 
We design this generative modeling framework to enable controllable and stable generation with a pretrained diffusion model \cite{song2020score, hess2025fast, ho2020denoising} through the incorporation of conditional guidance (Supplementary Notes 6 and 7).

\subsubsection*{\normalfont Diffusion model specification}
A diffusion model learns a reverse diffusion process that transforms a known Gaussian distribution \text{$\mathbf{Z}(t = T) \sim \mathcal{N}(0, I)$} into the target climatic data distribution \text{$\mathbf{Z}(t = 0) \approx \mathbf{X}$}, where $\mathbf{Z} \in \mathbb{R}^{n}$. In this framework, the forward diffusion process to progressively perturb the data follows the stochastic differential equation (SDE) \cite{song2020score,ho2020denoising}: 
\begin{equation}
\  d\mathbf{Z}
= \mu(\mathbf{Z},t)dt + g(t)dW,\quad t\in[0,T].
\end{equation}
Here $t$ denotes the step in the forward diffusion process. $W$ is a standard Wiener process. The drift term $\mu(\mathbf{Z},t)$ and the diffusion coefficient $g(t)$ are configured according to the setup of score-based generative modeling \cite{ho2020denoising, song2020score}. The reverse SDE used for generation is: 

\begin{equation}
d\mathbf{Z}
=
\Bigg[
\mu(\mathbf{Z},\overline{t})
- g^{2}(\overline{t}) \nabla_{\mathbf{Z}} \log p_{\overline{t}}
(\mathbf{Z} \mid \mathbf{X}_{d-1}, \mathbf{X}_{d-2})
\Bigg]
d\overline{t}
+ g(\overline{t}) d\overline{W}.
\end{equation}

$\overline{t}$ represents time reversal of the above SDE. $\nabla_{\mathbf{Z}}\log p_{\overline{t}}(\mathbf{Z}\mid \mathbf{X}_{d-1}, \mathbf{X}_{d-2})$ denotes the score function, where conditioning on $X_{d-1}$ and $X_{d-2}$ is introduced to enforce daily temporal consistency in the diffusion model \cite{dhariwal2021diffusion,price2025probabilistic}. 
Since this score function is not analytically tractable, we approximate it using a U-Net neural network,
$s_{\theta}(\mathbf{Z}, \overline{t}, \mathbf{X}_{d-1}, \mathbf{X}_{d-2}) \approx \nabla_{\mathbf{Z}}\log p_{\overline{t}}(\mathbf{Z}\mid \mathbf{X}_{d-1}, \mathbf{X}_{d-2})$, and train it to minimize the score matching loss through stochastic gradient descent \cite{song2020score}. 

\subsubsection*{\normalfont Neural network architectures and training}
The diffusion model is implemented using the standard Denoising Diffusion Probabilistic Model framework~\cite{ho2020denoising, song2020score}, conditioned on daily fields from the preceding two days. The score function $s_{\theta}$ is parameterized by a U-Net neural network \cite{song2020score}. A cosine noise schedule is employed to perturb the target climatic variables, while a linear noise schedule is applied to the conditional inputs during noise-conditioned data augmentation. The network was trained for 500 epochs using the Adam optimizer with a batch size of 1 and a learning rate of $1\times10^{-4}$, on an NVIDIA H100 GPU. During training, the reverse-time SDE in Eq.~(6) is integrated from $\overline{t}=0$ to $\overline{t}=T$ using the Euler--Maruyama scheme with 2000 steps. During inference, sampling is accelerated by reducing the number of steps to 200 using a second-order probability flow ordinary differential equation solver \cite{song2020denoising}, enabling fast generation while maintaining high-quality outputs. 

\subsubsection*{\normalfont Response pattern informed generative modeling}
In the RIG framework (Fig.~1), response theory produces the monthly response patterns $\mathbf{X}'_m$. Under the Bayesian formulation, they can serve as likelihood guidance ~\cite{chung2025diffusion} on the generation target $\mathbf{X}$ of the pretrained diffusion model. We adopt the linear conditional guidance approach~\cite{song2020score, chung2025diffusion} to estimate the likelihood-guidance score function $s_{g}(\mathbf{Z}, \mathbf{X}'_m)$ (see Supplementary Note 6 for details) which is then substituted into Eq.~(6). The conditional generation process is thus given by:
\begin{equation}
d\mathbf{Z}
=
\Bigg[
\mu(\mathbf{Z},\overline{t})
- g^{2}(\overline{t})
\left(
s_{\theta}(\mathbf{Z}, \overline{t}, \mathbf{X}_{d-1}, \mathbf{X}_{d-2})
+
s_{g}(\mathbf{Z}, \mathbf{X}'_m)
\right)
\Bigg]
d\overline{t}
+
g(\overline{t}) d\overline{W},
\end{equation}
where $s_{\theta}$ denotes the pretrained 
neural network that approximates the score function. This formulation provides a plug-and-play posterior sampler without additional retraining on the diffusion model~\cite{song2020score, hess2025fast, daras2024survey, chung2025diffusion} and enables RIG to control the daily climate fields generation conditioned on response theory-derived patterns.  

\bibliographystyle{sn-basic}
\bibliography{sn-bibliography}

\begin{thebibliography}{88}
\providecommand{\natexlab}[1]{#1}
\providecommand{\url}[1]{{#1}}
\providecommand{\urlprefix}{URL }
\providecommand{\doi}[1]{\url{https://doi.org/#1}}
\providecommand{\eprint}[2][]{\url{#2}}
 \bibcommenthead

\bibitem[{Aengenheyster et~al.(2018)Aengenheyster, Feng, Van Der~Ploeg, and Dijkstra}]{aengenheyster2018point}
Aengenheyster M, Feng QY, Van Der~Ploeg F, Dijkstra HA (2018) The point of no return for climate action: effects of climate uncertainty and risk tolerance. Earth System Dynamics 9(3):1085--1095

\bibitem[{Aich et~al.(2026)Aich, Hess, Pan, Bathiany, Huang, and Boers}]{aich2025conditional}
Aich M, Hess P, Pan B, Bathiany S, Huang Y, Boers N (2026) Conditional diffusion models for downscaling and bias correction of earth system model precipitation. Geoscientific Model Development 19(4):1791--1808. \doi{10.5194/gmd-19-1791-2026}

\bibitem[{Andrijevic et~al.(2025)Andrijevic, Zimm, Moyer, Muttarak, and Pachauri}]{andrijevic2025representing}
Andrijevic M, Zimm C, Moyer JD, Muttarak R, Pachauri S (2025) Representing gender inequality in scenarios improves understanding of climate challenges. Nature Climate Change 15(2):138--146

\bibitem[{Antonio et~al.(2026)Antonio, Strommen, and Christensen}]{antonio2026seasonal}
Antonio B, Strommen K, Christensen HM (2026) Seasonal forecasting using the gencast probabilistic machine learning model. Climate Dynamics 64(4):148

\bibitem[{Bathiany et~al.(2018)Bathiany, Dakos, Scheffer, and Lenton}]{bathiany2018climate}
Bathiany S, Dakos V, Scheffer M, Lenton TM (2018) Climate models predict increasing temperature variability in poor countries. Science advances 4(5):eaar5809

\bibitem[{Bevacqua et~al.(2026)Bevacqua, Fischer, Sillmann, and Zscheischler}]{bevacqua2026moderate}
Bevacqua E, Fischer E, Sillmann J, Zscheischler J (2026) Moderate global warming does not rule out extreme global climate outcomes. Nature 651(8107):946--953

\bibitem[{Boucher et~al.(2020)Boucher, Servonnat, Albright, Aumont, Balkanski, Bastrikov, Bekki, Bonnet, Bony, Bopp et~al.}]{boucher2020presentation}
Boucher O, Servonnat J, Albright AL, Aumont O, Balkanski Y, Bastrikov V, Bekki S, Bonnet R, Bony S, Bopp L, others (2020) Presentation and evaluation of the ipsl-cm6a-lr climate model. Journal of Advances in Modeling Earth Systems 12(7):e2019MS002010

\bibitem[{Brenowitz et~al.(2025)Brenowitz, Ge, Subramaniam, Manshausen, Gupta, Hall, Mardani, Vahdat, Kashinath, and Pritchard}]{brenowitz2025climate}
Brenowitz ND, Ge T, Subramaniam A, Manshausen P, Gupta A, Hall DM, Mardani M, Vahdat A, Kashinath K, Pritchard MS (2025) Climate in a bottle: Towards a generative foundation model for the kilometer-scale global atmosphere. arXiv preprint arXiv:250506474

\bibitem[{Buster et~al.(2024)Buster, Benton, Glaws, and King}]{buster2024high}
Buster G, Benton BN, Glaws A, King RN (2024) High-resolution meteorology with climate change impacts from global climate model data using generative machine learning. Nature Energy 9(7):894--906

\bibitem[{Cannon(2018)}]{cannon2018multivariate}
Cannon AJ (2018) Multivariate quantile mapping bias correction: an n-dimensional probability density function transform for climate model simulations of multiple variables. Climate dynamics 50(1):31--49

\bibitem[{Cannon et~al.(2015)Cannon, Sobie, and Murdock}]{cannon2015bias}
Cannon AJ, Sobie SR, Murdock TQ (2015) Bias correction of gcm precipitation by quantile mapping: how well do methods preserve changes in quantiles and extremes? Journal of Climate 28(17):6938--6959

\bibitem[{Chen et~al.(2024)Chen, Mart{\'\i}~Mons{\'o}, Du, Simchowitz, Tedrake, and Sitzmann}]{chen2024diffusion}
Chen B, Mart{\'\i}~Mons{\'o} D, Du Y, Simchowitz M, Tedrake R, Sitzmann V (2024) Diffusion forcing: Next-token prediction meets full-sequence diffusion. Advances in Neural Information Processing Systems 37:24081--24125

\bibitem[{Chung et~al.(2022)Chung, Kim, Mccann, Klasky, and Ye}]{chung2022diffusion}
Chung H, Kim J, Mccann MT, Klasky ML, Ye JC (2022) Diffusion posterior sampling for general noisy inverse problems. arXiv preprint arXiv:220914687

\bibitem[{Chung et~al.(2025)Chung, Kim, and Ye}]{chung2025diffusion}
Chung H, Kim J, Ye JC (2025) Diffusion models for inverse problems. arXiv preprint arXiv:250801975

\bibitem[{Craig et~al.(2022)Craig, Wohland, Stoop, Kies, Pickering, Bloomfield, Browell, De~Felice, Dent, Deroubaix et~al.}]{craig2022overcoming}
Craig MT, Wohland J, Stoop LP, Kies A, Pickering B, Bloomfield HC, Browell J, De~Felice M, Dent CJ, Deroubaix A, others (2022) Overcoming the disconnect between energy system and climate modeling. Joule 6(7):1405--1417

\bibitem[{Creswell et~al.(2018)Creswell, White, Dumoulin, Arulkumaran, Sengupta, and Bharath}]{creswell2018generative}
Creswell A, White T, Dumoulin V, Arulkumaran K, Sengupta B, Bharath AA (2018) Generative adversarial networks: An overview. IEEE signal processing magazine 35(1):53--65

\bibitem[{Danabasoglu et~al.(2020)Danabasoglu, Lamarque, Bacmeister, Bailey, DuVivier, Edwards, Emmons, Fasullo, Garcia, Gettelman et~al.}]{danabasoglu2020community}
Danabasoglu G, Lamarque JF, Bacmeister J, Bailey D, DuVivier A, Edwards J, Emmons L, Fasullo J, Garcia R, Gettelman A, others (2020) The community earth system model version 2 (cesm2). Journal of Advances in Modeling Earth Systems 12(2):e2019MS001916

\bibitem[{Daras et~al.(2024)Daras, Chung, Lai, Mitsufuji, Ye, Milanfar, Dimakis, and Delbracio}]{daras2024survey}
Daras G, Chung H, Lai CH, Mitsufuji Y, Ye JC, Milanfar P, Dimakis AG, Delbracio M (2024) A survey on diffusion models for inverse problems. arXiv preprint arXiv:241000083

\bibitem[{Dentener et~al.(2021)Dentener, Hall, Smith et~al.}]{dentener2021annex}
Dentener F, Hall B, Smith C, others (2021) Annex iii: Tables of historical and projected well-mixed greenhouse gas mixing ratios and effective radiative forcing of all climate forcers

\bibitem[{Dhariwal and Nichol(2021)}]{dhariwal2021diffusion}
Dhariwal P, Nichol A (2021) Diffusion models beat gans on image synthesis. Advances in neural information processing systems 34:8780--8794

\bibitem[{Di~Virgilio et~al.(2025)Di~Virgilio, Ji, Tam, Evans, Kala, Andrys, Thomas, Choudhury, Rocha, Li et~al.}]{di2025evaluation}
Di~Virgilio G, Ji F, Tam E, Evans JP, Kala J, Andrys J, Thomas C, Choudhury D, Rocha C, Li Y, others (2025) Evaluation of cordex era5-forced narclim2. 0 regional climate models over australia using the weather research and forecasting (wrf) model version 4.1. 2. Geoscientific Model Development 18(3):703--724

\bibitem[{Dinh and Aires(2023)}]{dinh2023revisiting}
Dinh TLA, Aires F (2023) Revisiting the bias correction of climate models for impact studies. Climatic Change 176(10):140

\bibitem[{Doblas-Reyes et~al.(2021)Doblas-Reyes, S{\"o}rensson, Almazroui, Dosio, Gutowski, Haarsma, Hamdi, Hewitson, Kwon, Lamptey, Maraun, Stephenson, Takayabu, Terray, Turner, and Zuo}]{DoblasReyes2021}
Doblas-Reyes FJ, S{\"o}rensson AA, Almazroui M, Dosio A, Gutowski WJ, Haarsma R, Hamdi R, Hewitson B, Kwon WT, Lamptey BL, Maraun D, Stephenson TS, Takayabu I, Terray L, Turner A, Zuo Z (2021) Linking global to regional climate change. In: Masson-Delmotte V, Zhai P, Pirani A, Connors SL, P{\'e}an C, Berger S, Caud N, Chen Y, Goldfarb L, Gomis MI, Huang M, Leitzell K, Lonnoy E, Matthews JBR, Maycock TK, Waterfield T, Yelek{\c{c}}i O, Yu R, Zhou B (eds) Climate Change 2021: The Physical Science Basis. Cambridge University Press, Cambridge, United Kingdom and New York, NY, USA, p 1363--1512, \doi{10.1017/9781009157896.012}

\bibitem[{Easterling et~al.(2000)Easterling, Meehl, Parmesan, Changnon, Karl, and Mearns}]{easterling2000climate}
Easterling DR, Meehl GA, Parmesan C, Changnon SA, Karl TR, Mearns LO (2000) Climate extremes: observations, modeling, and impacts. science 289(5487):2068--2074

\bibitem[{Eyring et~al.(2016)Eyring, Bony, Meehl, Senior, Stevens, Stouffer, and Taylor}]{eyring2016overview}
Eyring V, Bony S, Meehl GA, Senior CA, Stevens B, Stouffer RJ, Taylor KE (2016) Overview of the coupled model intercomparison project phase 6 (cmip6) experimental design and organization. Geoscientific Model Development 9(5):1937--1958

\bibitem[{Eyring et~al.(2023)Eyring, Gillett, Achuta~Rao, Barimalala, Barreiro~Parrillo, Bellouin, Cassou, Durack, Kosaka, McGregor et~al.}]{eyring2023human}
Eyring V, Gillett NP, Achuta~Rao KM, Barimalala R, Barreiro~Parrillo M, Bellouin N, Cassou C, Durack PJ, Kosaka Y, McGregor S, others (2023) Human influence on the climate system (chapter 3). IPCC 2021: Climate Change 2021: The Physical Science Basis Contribution of Working Group I to the Sixth Assessment Report of the Intergovernmental Panel on Climate Change pp 423--552

\bibitem[{Fischer et~al.(2021)Fischer, Sippel, and Knutti}]{fischer2021increasing}
Fischer EM, Sippel S, Knutti R (2021) Increasing probability of record-shattering climate extremes. Nature Climate Change 11(8):689--695

\bibitem[{Fischer et~al.(2023)Fischer, Beyerle, Bloin-Wibe, Gessner, Humphrey, Lehner, Pendergrass, Sippel, Zeder, and Knutti}]{fischer2023storylines}
Fischer EM, Beyerle U, Bloin-Wibe L, Gessner C, Humphrey V, Lehner F, Pendergrass AG, Sippel S, Zeder J, Knutti R (2023) Storylines for unprecedented heatwaves based on ensemble boosting. Nature Communications 14(1):4643

\bibitem[{Gelbrecht et~al.(2023)Gelbrecht, White, Bathiany, and Boers}]{gelbrecht2023differentiable}
Gelbrecht M, White A, Bathiany S, Boers N (2023) Differentiable programming for earth system modeling. Geoscientific Model Development 16(11):3123--3135

\bibitem[{Gergel et~al.(2024)Gergel, Malevich, McCusker, Tenezakis, Delgado, Fish, and Kopp}]{gergel2024global}
Gergel DR, Malevich SB, McCusker KE, Tenezakis E, Delgado MT, Fish MA, Kopp RE (2024) Global downscaled projections for climate impacts research (gdpcir): Preserving quantile trends for modeling future climate impacts. Geoscientific Model Development 17(1):191--227

\bibitem[{Giorgini et~al.(2024)Giorgini, Deck, Bischoff, and Souza}]{giorgini2024response}
Giorgini LT, Deck K, Bischoff T, Souza A (2024) Response theory via generative score modeling. Physical Review Letters 133(26):267302

\bibitem[{Giorgini et~al.(2025)Giorgini, Falasca, and Souza}]{giorgini2025predicting}
Giorgini LT, Falasca F, Souza AN (2025) Predicting forced responses of probability distributions via the fluctuation--dissipation theorem and generative modeling. Proceedings of the National Academy of Sciences 122(41):e2509578122

\bibitem[{Haywood et~al.(2016)Haywood, Dowsett, Dolan, Rowley, Abe-Ouchi, Otto-Bliesner, Chandler, Hunter, Lunt, Pound et~al.}]{haywood2016pliocene}
Haywood AM, Dowsett HJ, Dolan AM, Rowley D, Abe-Ouchi A, Otto-Bliesner B, Chandler MA, Hunter SJ, Lunt DJ, Pound M, others (2016) The pliocene model intercomparison project (pliomip) phase 2: scientific objectives and experimental design. Climate of the Past 12(3):663--675

\bibitem[{Hegerl et~al.(1996)Hegerl, voN SToRcH, Hasselmann, Santer, Cubasch, and Jones}]{hegerl1996detecting}
Hegerl GC, voN SToRcH H, Hasselmann K, Santer BD, Cubasch U, Jones PD (1996) Detecting greenhouse-gas-induced climate change with an optimal fingerprint method. Journal of Climate 9(10):2281--2306

\bibitem[{Hersbach et~al.(2020)Hersbach, Bell, Berrisford, Hirahara, Hor{\'a}nyi, Mu{\~n}oz-Sabater, Nicolas, Peubey, Radu, Schepers et~al.}]{hersbach2020era5}
Hersbach H, Bell B, Berrisford P, Hirahara S, Hor{\'a}nyi A, Mu{\~n}oz-Sabater J, Nicolas J, Peubey C, Radu R, Schepers D, others (2020) The era5 global reanalysis. Quarterly journal of the royal meteorological society 146(730):1999--2049

\bibitem[{Hess et~al.(2022)Hess, Dr{\"u}ke, Petri, Strnad, and Boers}]{hess2022physically}
Hess P, Dr{\"u}ke M, Petri S, Strnad FM, Boers N (2022) Physically constrained generative adversarial networks for improving precipitation fields from earth system models. Nature Machine Intelligence 4(10):828--839

\bibitem[{Hess et~al.(2025)Hess, Aich, Pan, and Boers}]{hess2025fast}
Hess P, Aich M, Pan B, Boers N (2025) Fast, scale-adaptive and uncertainty-aware downscaling of earth system model fields with generative machine learning. Nature Machine Intelligence pp 1--11

\bibitem[{Ho et~al.(2020)Ho, Jain, and Abbeel}]{ho2020denoising}
Ho J, Jain A, Abbeel P (2020) Denoising diffusion probabilistic models. Advances in neural information processing systems 33:6840--6851

\bibitem[{Huang et~al.(2026)Huang, Yang, Bathiany, Aich, Hess, and Boers}]{huang2026generative}
Huang Y, Yang S, Bathiany S, Aich M, Hess P, Boers N (2026) Generative modeling for multivariate spatio-temporal downscaling of monthly climate fields. Manuscript to be submitted (see Supplementary Material-2)

\bibitem[{Huisman et~al.(2025)Huisman, Martyr, Rott, and Smits}]{huisman2025projections}
Huisman J, Martyr R, Rott R, Smits J (2025) Projections of climate change vulnerability along the shared socioeconomic pathways 2020--2100. Scientific Data 12(1):1527

\bibitem[{{Intergovernmental Panel on Climate Change}(2022)}]{IPCC2022WGII_Full}
{Intergovernmental Panel on Climate Change} (2022) Climate Change 2022: Impacts, Adaptation and Vulnerability. Contribution of Working Group II to the Sixth Assessment Report of the Intergovernmental Panel on Climate Change. Cambridge University Press, Cambridge, UK and New York, NY, USA, \doi{10.1017/9781009325844}

\bibitem[{Knutti et~al.(2017)Knutti, Rugenstein, and Hegerl}]{knutti2017beyond}
Knutti R, Rugenstein MA, Hegerl GC (2017) Beyond equilibrium climate sensitivity. Nature Geoscience 10(10):727--736

\bibitem[{Kochkov et~al.(2024)Kochkov, Yuval, Langmore, Norgaard, Smith, Mooers, Kl{\"o}wer, Lottes, Rasp, D{\"u}ben et~al.}]{kochkov2024neural}
Kochkov D, Yuval J, Langmore I, Norgaard P, Smith J, Mooers G, Kl{\"o}wer M, Lottes J, Rasp S, D{\"u}ben P, others (2024) Neural general circulation models for weather and climate. Nature 632(8027):1060--1066

\bibitem[{Lee et~al.(2021)Lee, Marotzke, Bala, Cao, Corti, Dunne, Engelbrecht, Fischer, Fyfe, Jones, Maycock, Mutemi, Ndiaye, Panickal, and Zhou}]{IPCC_AR6_WGI_Chapter4_2021}
Lee JY, Marotzke J, Bala G, Cao L, Corti S, Dunne J, Engelbrecht F, Fischer E, Fyfe J, Jones C, Maycock A, Mutemi J, Ndiaye O, Panickal S, Zhou T (2021) Future Global Climate: Scenario-Based Projections and Near-Term Information, Cambridge University Press, Cambridge, United Kingdom and New York, NY, USA, pp 553--672. \doi{10.1017/9781009157896.006}

\bibitem[{Lee et~al.(2025)Lee, Sharma, Rosenbloom, Rodgers, Kim, Kwon, Franzke, Kim, Sreeush, and Stein}]{beyond2100}
Lee SS, Sharma S, Rosenbloom N, Rodgers KB, Kim JE, Kwon EY, Franzke CLE, Kim IW, Sreeush MG, Stein K (2025) Multi-centennial climate change in a warming world beyond 2100. Earth System Dynamics 16(5):1427--1451

\bibitem[{Lembo et~al.(2020)Lembo, Lucarini, and Ragone}]{lembo2020beyond}
Lembo V, Lucarini V, Ragone F (2020) Beyond forcing scenarios: predicting climate change through response operators in a coupled general circulation model. Scientific Reports 10(1):8668

\bibitem[{Li and He(2025)}]{li2025back}
Li T, He K (2025) Back to basics: Let denoising generative models denoise. arXiv preprint arXiv:251113720

\bibitem[{Lipman et~al.(2022)Lipman, Chen, Ben-Hamu, Nickel, and Le}]{lipman2022flow}
Lipman Y, Chen RT, Ben-Hamu H, Nickel M, Le M (2022) Flow matching for generative modeling. arXiv preprint arXiv:221002747

\bibitem[{Lopez-Gomez et~al.(2025)Lopez-Gomez, Wan, Zepeda-N{\'u}{\~n}ez, Schneider, Anderson, and Sha}]{lopez2025dynamical}
Lopez-Gomez I, Wan ZY, Zepeda-N{\'u}{\~n}ez L, Schneider T, Anderson J, Sha F (2025) Dynamical-generative downscaling of climate model ensembles. Proceedings of the National Academy of Sciences 122(17):e2420288122

\bibitem[{Lucarini and Chekroun(2023)}]{lucarini2023theoretical}
Lucarini V, Chekroun MD (2023) Theoretical tools for understanding the climate crisis from hasselmann’s programme and beyond. Nature Reviews Physics 5(12):744--765

\bibitem[{Lucarini and Chekroun(2024)}]{lucarini2024detecting}
Lucarini V, Chekroun MD (2024) Detecting and attributing change in climate and complex systems: Foundations, green’s functions, and nonlinear fingerprints. Physical Review Letters 133(24):244201

\bibitem[{Lucarini et~al.(2017)Lucarini, Ragone, and Lunkeit}]{lucarini2017predicting}
Lucarini V, Ragone F, Lunkeit F (2017) Predicting climate change using response theory: Global averages and spatial patterns. Journal of Statistical Physics 166(3):1036--1064

\bibitem[{Ma et~al.(2026)Ma, Abraham, Versick, Ruhnke, Schneidereit, Niemeier, Back, Braesicke, and Nowack}]{ma2026mloz}
Ma Y, Abraham NL, Versick S, Ruhnke R, Schneidereit A, Niemeier U, Back F, Braesicke P, Nowack P (2026) mloz: A highly efficient machine learning-based ozone parameterization for climate sensitivity simulations. Journal of Advances in Modeling Earth Systems 18(4):e2025MS005459

\bibitem[{Manshausen et~al.(2025)Manshausen, Cohen, Harrington, Pathak, Pritchard, Garg, Mardani, Kashinath, Byrne, and Brenowitz}]{manshausen2025generative}
Manshausen P, Cohen Y, Harrington P, Pathak J, Pritchard M, Garg P, Mardani M, Kashinath K, Byrne S, Brenowitz N (2025) Generative data assimilation of sparse weather station observations at kilometer scales. Journal of Advances in Modeling Earth Systems 17(10):e2024MS004505

\bibitem[{Maraun and Widmann(2018)}]{maraun2018statistical}
Maraun D, Widmann M (2018) Statistical downscaling and bias correction for climate research. Cambridge University Press

\bibitem[{Masson-Delmotte et~al.(2021)Masson-Delmotte, Zhai, Pirani, Connors, P{\'e}an, Berger, Caud, Chen, Goldfarb, Gomis et~al.}]{masson2021climate}
Masson-Delmotte V, Zhai P, Pirani A, Connors SL, P{\'e}an C, Berger S, Caud N, Chen Y, Goldfarb L, Gomis MI, others (2021) Climate change 2021: the physical science basis. Contribution of working group I to the sixth assessment report of the intergovernmental panel on climate change 2(1):2391

\bibitem[{Mauritsen et~al.(2019)Mauritsen, Bader, Becker, Behrens, Bittner, Brokopf, Brovkin, Claussen, Crueger, Esch et~al.}]{mauritsen2019developments}
Mauritsen T, Bader J, Becker T, Behrens J, Bittner M, Brokopf R, Brovkin V, Claussen M, Crueger T, Esch M, others (2019) Developments in the mpi-m earth system model version 1.2 (mpi-esm1. 2) and its response to increasing co2. Journal of Advances in Modeling Earth Systems 11(4):998--1038

\bibitem[{Meehl et~al.(2020)Meehl, Senior, Eyring, Flato, Lamarque, Stouffer, Taylor, and Schlund}]{meehl2020context}
Meehl GA, Senior CA, Eyring V, Flato G, Lamarque JF, Stouffer RJ, Taylor KE, Schlund M (2020) Context for interpreting equilibrium climate sensitivity and transient climate response from the cmip6 earth system models. Science Advances 6(26):eaba1981

\bibitem[{Meinshausen et~al.(2020)Meinshausen, Nicholls, Lewis, Gidden, Vogel, Freund, Beyerle, Gessner, Nauels, Bauer et~al.}]{meinshausen2020shared}
Meinshausen M, Nicholls ZR, Lewis J, Gidden MJ, Vogel E, Freund M, Beyerle U, Gessner C, Nauels A, Bauer N, others (2020) The shared socio-economic pathway (ssp) greenhouse gas concentrations and their extensions to 2500. Geoscientific Model Development 13(8):3571--3605

\bibitem[{Moses et~al.(2025)Moses, Cheng, Churavy, Gelbrecht, Kl{\"o}wer, Kump, Morlighem, Williamson, Apte, Berg et~al.}]{moses2025dj4earth}
Moses WS, Cheng G, Churavy V, Gelbrecht M, Kl{\"o}wer M, Kump J, Morlighem M, Williamson S, Apte D, Berg P, others (2025) Dj4earth: Differentiable, and performance-portable earth system modeling via program transformations. Authorea Preprints

\bibitem[{Nai et~al.(2025)Nai, Chen, Yang, Xiao, and Pan}]{nai2025boosting}
Nai C, Chen X, Yang S, Xiao Z, Pan B (2025) Boosting weather forecast via generative superensemble. npj Climate and Atmospheric Science 8(1):377

\bibitem[{O'Neill et~al.(2016)O'Neill, Tebaldi, Van~Vuuren, Eyring, Friedlingstein, Hurtt, Knutti, Kriegler, Lamarque, Lowe et~al.}]{o2016scenario}
O'Neill BC, Tebaldi C, Van~Vuuren DP, Eyring V, Friedlingstein P, Hurtt G, Knutti R, Kriegler E, Lamarque JF, Lowe J, others (2016) The scenario model intercomparison project (scenariomip) for cmip6. Geoscientific Model Development 9(9):3461--3482

\bibitem[{O’Neill et~al.(2014)O’Neill, Kriegler, Riahi, Ebi, Hallegatte, Carter, Mathur, and Van~Vuuren}]{o2014new}
O’Neill BC, Kriegler E, Riahi K, Ebi KL, Hallegatte S, Carter TR, Mathur R, Van~Vuuren DP (2014) A new scenario framework for climate change research: the concept of shared socioeconomic pathways. Climatic change 122(3):387--400

\bibitem[{Palmer and Stevens(2019)}]{palmer2019scientific}
Palmer T, Stevens B (2019) The scientific challenge of understanding and estimating climate change. Proceedings of the National Academy of Sciences 116(49):24390--24395

\bibitem[{Palmer(2019)}]{palmer2019stochastic}
Palmer TN (2019) Stochastic weather and climate models. Nature Reviews Physics 1(7):463--471

\bibitem[{Price et~al.(2025)Price, Sanchez-Gonzalez, Alet, Andersson, El-Kadi, Masters, Ewalds, Stott, Mohamed, Battaglia et~al.}]{price2025probabilistic}
Price I, Sanchez-Gonzalez A, Alet F, Andersson TR, El-Kadi A, Masters D, Ewalds T, Stott J, Mohamed S, Battaglia P, others (2025) Probabilistic weather forecasting with machine learning. Nature 637(8044):84--90

\bibitem[{Pörtner et~al.(2022)Pörtner, Roberts, Adams, Adelekan, Adler, Adrian, Aldunce, Ali, Begum, Friedl, Kerr, Biesbroek, Birkmann, Bowen, Caretta, Carnicer, Castellanos, Cheong, Chow, G.~Cissé, and Ibrahim}]{RN23}
Pörtner HO, Roberts D, Adams H, Adelekan I, Adler C, Adrian R, Aldunce P, Ali E, Begum RA, Friedl BB, Kerr RB, Biesbroek R, Birkmann J, Bowen K, Caretta M, Carnicer J, Castellanos E, Cheong T, Chow W, G.~Cissé GC, Ibrahim ZZ (2022) Climate Change 2022: Impacts, Adaptation and Vulnerability. Technical Summary, Cambridge University Press, Cambridge, UK and New York, USA

\bibitem[{Qian and Chang(2021)}]{qian2021projecting}
Qian W, Chang HH (2021) Projecting health impacts of future temperature: a comparison of quantile-mapping bias-correction methods. International journal of environmental research and public health 18(4):1992

\bibitem[{Ragone et~al.(2016)Ragone, Lucarini, and Lunkeit}]{ragone2016new}
Ragone F, Lucarini V, Lunkeit F (2016) A new framework for climate sensitivity and prediction: a modelling perspective. Climate dynamics 46(5):1459--1471

\bibitem[{Rodgers et~al.(2021)Rodgers, Lee, Rosenbloom, Timmermann, Danabasoglu, Deser, Edwards, Kim, Simpson, Stein et~al.}]{rodgers2021ubiquity}
Rodgers KB, Lee SS, Rosenbloom N, Timmermann A, Danabasoglu G, Deser C, Edwards J, Kim JE, Simpson IR, Stein K, others (2021) Ubiquity of human-induced changes in climate variability. Earth System Dynamics 12(4):1393--1411

\bibitem[{Ruane et~al.(2015)Ruane, Goldberg, and Chryssanthacopoulos}]{ruane2015climate}
Ruane AC, Goldberg R, Chryssanthacopoulos J (2015) Climate forcing datasets for agricultural modeling: Merged products for gap-filling and historical climate series estimation. Agricultural and Forest Meteorology 200:233--248

\bibitem[{Ruelle(1998)}]{ruelle1998general}
Ruelle D (1998) General linear response formula in statistical mechanics, and the fluctuation-dissipation theorem far from equilibrium. Physics Letters A 245(3-4):220--224

\bibitem[{Ruelle(2009)}]{ruelle2009review}
Ruelle D (2009) A review of linear response theory for general differentiable dynamical systems. Nonlinearity 22(4):855

\bibitem[{Rugenstein et~al.(2020)Rugenstein, Bloch-Johnson, Gregory, Andrews, Mauritsen, Li, Fr{\"o}licher, Paynter, Danabasoglu, Yang et~al.}]{rugenstein2020equilibrium}
Rugenstein M, Bloch-Johnson J, Gregory J, Andrews T, Mauritsen T, Li C, Fr{\"o}licher TL, Paynter D, Danabasoglu G, Yang S, others (2020) Equilibrium climate sensitivity estimated by equilibrating climate models. Geophysical Research Letters 47(4):e2019GL083898

\bibitem[{Sandstad et~al.(2025)Sandstad, Steinert, Baur, and Sanderson}]{sandstad2025meteorv1}
Sandstad M, Steinert NJ, Baur S, Sanderson BM (2025) Meteorv1. 0.1: A novel framework for emulating multi-timescale regional climate responses. Geoscientific Model Development 18(21):8269--8312

\bibitem[{Sippel et~al.(2020)Sippel, Meinshausen, Fischer, Sz{\'e}kely, and Knutti}]{sippel2020climate}
Sippel S, Meinshausen N, Fischer EM, Sz{\'e}kely E, Knutti R (2020) Climate change now detectable from any single day of weather at global scale. Nature climate change 10(1):35--41

\bibitem[{Smith et~al.(2021)Smith, Hall, Dentener, Ahn, Collins, Jones, Meinshausen, Dlugokencky, Keeling, Krummel, M{\"u}hle, Nicholls, and Simpson}]{Smith2021IPCC_AR6_AnnexIII}
Smith C, Hall B, Dentener F, Ahn J, Collins W, Jones C, Meinshausen M, Dlugokencky E, Keeling R, Krummel P, M{\"u}hle J, Nicholls Z, Simpson I (2021) {IPCC Working Group 1 (WG1) Sixth Assessment Report (AR6) Annex III Extended Data}. \doi{10.5281/zenodo.5705391}, \urlprefix\url{https://doi.org/10.5281/zenodo.5705391}

\bibitem[{Song et~al.(2020{\natexlab{a}})Song, Meng, and Ermon}]{song2020denoising}
Song J, Meng C, Ermon S (2020{\natexlab{a}}) Denoising diffusion implicit models. arXiv preprint arXiv:201002502

\bibitem[{Song et~al.(2020{\natexlab{b}})Song, Sohl-Dickstein, Kingma, Kumar, Ermon, and Poole}]{song2020score}
Song Y, Sohl-Dickstein J, Kingma DP, Kumar A, Ermon S, Poole B (2020{\natexlab{b}}) Score-based generative modeling through stochastic differential equations. arXiv preprint arXiv:201113456

\bibitem[{Sun et~al.(2018)Sun, Miao, Duan, Ashouri, Sorooshian, and Hsu}]{sun2018review}
Sun Q, Miao C, Duan Q, Ashouri H, Sorooshian S, Hsu KL (2018) A review of global precipitation data sets: Data sources, estimation, and intercomparisons. Reviews of geophysics 56(1):79--107

\bibitem[{Sun et~al.(2025)Sun, Nai, Pan, Li, Li, Li, Foufoula-Georgiou, and Lin}]{sun2025fusion}
Sun S, Nai C, Pan B, Li W, Li L, Li X, Foufoula-Georgiou E, Lin Y (2025) Fusion of multi-source precipitation records via coordinate-based generative models. Nature Communications

\bibitem[{Tebaldi et~al.(2021)Tebaldi, Debeire, Eyring, Fischer, Fyfe, Friedlingstein, Knutti, Lowe, O'Neill, Sanderson et~al.}]{tebaldi2021climate}
Tebaldi C, Debeire K, Eyring V, Fischer E, Fyfe J, Friedlingstein P, Knutti R, Lowe J, O'Neill B, Sanderson B, others (2021) Climate model projections from the scenario model intercomparison project (scenariomip) of cmip6. Earth System Dynamics 12(1):253--293

\bibitem[{Terhaar et~al.(2022)Terhaar, Fr{\"o}licher, Aschwanden, Friedlingstein, and Joos}]{terhaar2022adaptive}
Terhaar J, Fr{\"o}licher TL, Aschwanden MT, Friedlingstein P, Joos F (2022) Adaptive emission reduction approach to reach any global warming target. Nature Climate Change 12(12):1136--1142

\bibitem[{van Vuuren et~al.(2025)van Vuuren, Doelman, Schmidt~Tagomori, Beusen, Cornell, R{\"o}ckstrom, Schipper, Stehfest, Ambrosio, van~den Berg et~al.}]{van2025exploring}
van Vuuren DP, Doelman JC, Schmidt~Tagomori I, Beusen AH, Cornell SE, R{\"o}ckstrom J, Schipper AM, Stehfest E, Ambrosio G, van~den Berg M, others (2025) Exploring pathways for world development within planetary boundaries. Nature pp 1--7

\bibitem[{Watt-Meyer et~al.(2025)Watt-Meyer, Henn, McGibbon, Clark, Kwa, Perkins, Wu, Harris, and Bretherton}]{watt2025ace2}
Watt-Meyer O, Henn B, McGibbon J, Clark SK, Kwa A, Perkins WA, Wu E, Harris L, Bretherton CS (2025) Ace2: accurately learning subseasonal to decadal atmospheric variability and forced responses. npj Climate and Atmospheric Science 8(1):205

\bibitem[{Womack et~al.(2025)Womack, Giani, Eastham, and Selin}]{womack2025rapid}
Womack CB, Giani P, Eastham SD, Selin NE (2025) Rapid emulation of spatially resolved temperature response to effective radiative forcing. Journal of Advances in Modeling Earth Systems 17(1):e2024MS004523

\bibitem[{Yang et~al.(2025)Yang, Nai, Liu, Li, Chao, Wang, Wang, Li, Chen, Lu et~al.}]{yang2025generative}
Yang S, Nai C, Liu X, Li W, Chao J, Wang J, Wang L, Li X, Chen X, Lu B, others (2025) Generative assimilation and prediction for weather and climate. arXiv preprint arXiv:250303038

\bibitem[{Yoon et~al.(2026)Yoon, Hohenegger, Bao, and Brunner}]{yoon2026extreme}
Yoon A, Hohenegger C, Bao J, Brunner L (2026) Extreme events in the amazon after deforestation. Earth System Dynamics 17(1):167--179

\end{thebibliography}

\bmhead{Data availability}
The surface air temperature and total precipitation data from the MPI-ESM1.2-LR, CESM2 and IPSL-CM6A-LR simulations are archived in the CMIP6 by the Earth System Grid Federation (ESGF): \url{https://esgf-metagrid.cloud.dkrz.de/search/cmip6}. The reanalysis data for training diffusion model were obtained from Copernicus Climate Data Store for ERA5 reanalysis: \url{https://cds.climate.copernicus.eu/datasets/reanalysis-era5-single-levels}. The effective radiative forcings for all climate forcing components across different SSPs are provided by Intergovernmental Panel on Climate Change
(IPCC) Working Group I \cite{dentener2021annex, Smith2021IPCC_AR6_AnnexIII}: \url{https://doi.org/10.5281/zenodo.5705391}

\bmhead{Code availability}
The diffusion model used in this study was implemented in Python~3.12 using PyTorch~2.2.0. All analyses were conducted in Python~3.12. The source code for response theory and generative machine learning will be made publicly available upon publication via Zenodo and GitHub. The computation of spatial power spectrum was based on the torch-harmonic python library at \url{https://github.com/NVIDIA/torch-harmonics}.

\bmhead{Acknowledgements}
We acknowledge the ClimTip project that has received funding from the European Union’s Horizon Europe research and innovation programme under grant agreement No. 101137601; This is ClimTip contribution \# 161. The authors gratefully acknowledge the Ministry of Research, Science and Culture (MWFK) of Land Brandenburg for supporting this project by providing resources on the high performance computer system at the Potsdam Institute for Climate Impact Research. 
Y.H. also acknowledges Alexander von Humboldt Foundation for Humboldt Research fellowship. N.B. also acknowledges further support by the Volkswagen foundation, the Deutsche Forschungsgemeinschaft (DFG), the National Research Foundation of Korea (NRF): Climate Resilience under Zero-Emission Commitment Scenarios, BO 4455/2-1 (eBer-24-62950; RS-2024-00438471), the advanced Research and Invention Agency (ARIA), and the Past to Future (P2F) project, which has received funding from the European Union’s Horizon Europe research and innovation programme under grant agreement No. 101184070.

\bmhead{Author contributions}
Y.H., N.B., and S.B. conceived and designed the research. Y.H. and S.Y. improved the generative machine learning algorithms, with contributions from P.H., M.A., and N.B. Y.H. improved the response theory algorithms, with contributions from S.B. and N.B. Y.H. performed the programming, prepared the data, and conducted the numerical analyses. All authors contributed to the discussion, interpretation, and validation of the results. Y.H. carried out the visualization and wrote the manuscript with input from all authors.

\bmhead{Competing interests}
The authors declare no competing interests.

\clearpage
\begin{figure}[htbp]
\centering
\includegraphics[width=1.05\textwidth]{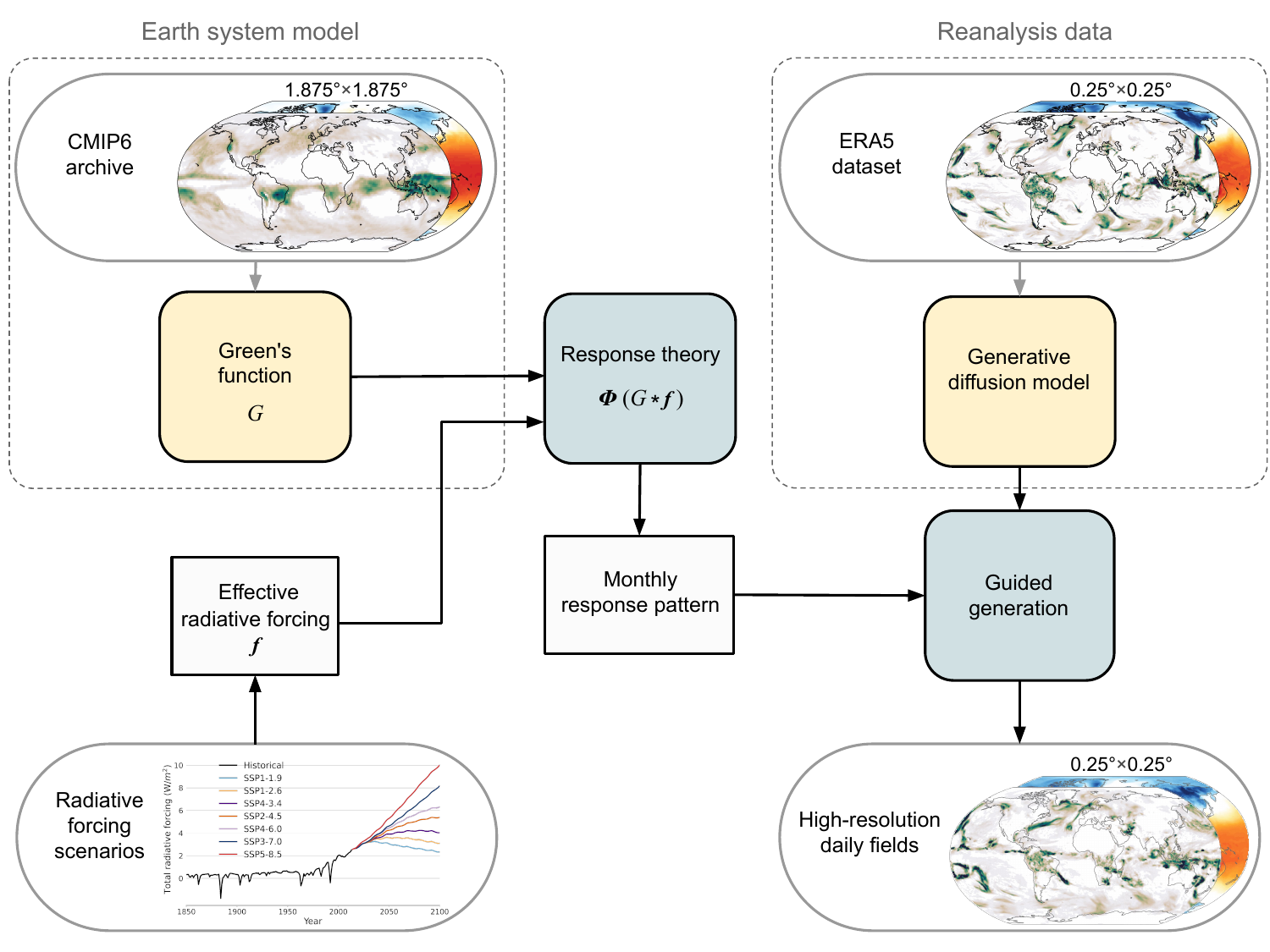}
\caption{\textbf{Schematic of the RIG framework for climate fields projection under a given radiative forcing scenario.} In the training stage, the Green's function $G$ is estimated using temperature and precipitation data from the abrupt CO$_2$ quadrupling experiment of an Earth system model from CMIP6. Concurrently, a generative diffusion model is trained on ERA5 reanalysis data to generate time- and variable-consistent sequences high-resolution daily global temperature and precipitation fields, while learning the joint probabilistic structure across daily and monthly climate fields. In the inference stage, given the effective radiative forcing of an arbitrary SSP scenario, the response theory model $\mathbf{\Phi}$ predicts monthly coarse-resolution responses of temperature and precipitation, which then act as conditional guidance (Methods) for the diffusion model to generate consecutive, bias corrected high-resolution daily temperature and precipitation fields.  }\label{fig1}
\end{figure}

\clearpage
\begin{figure}[htbp]
\centering
\includegraphics[width=1.01\textwidth]{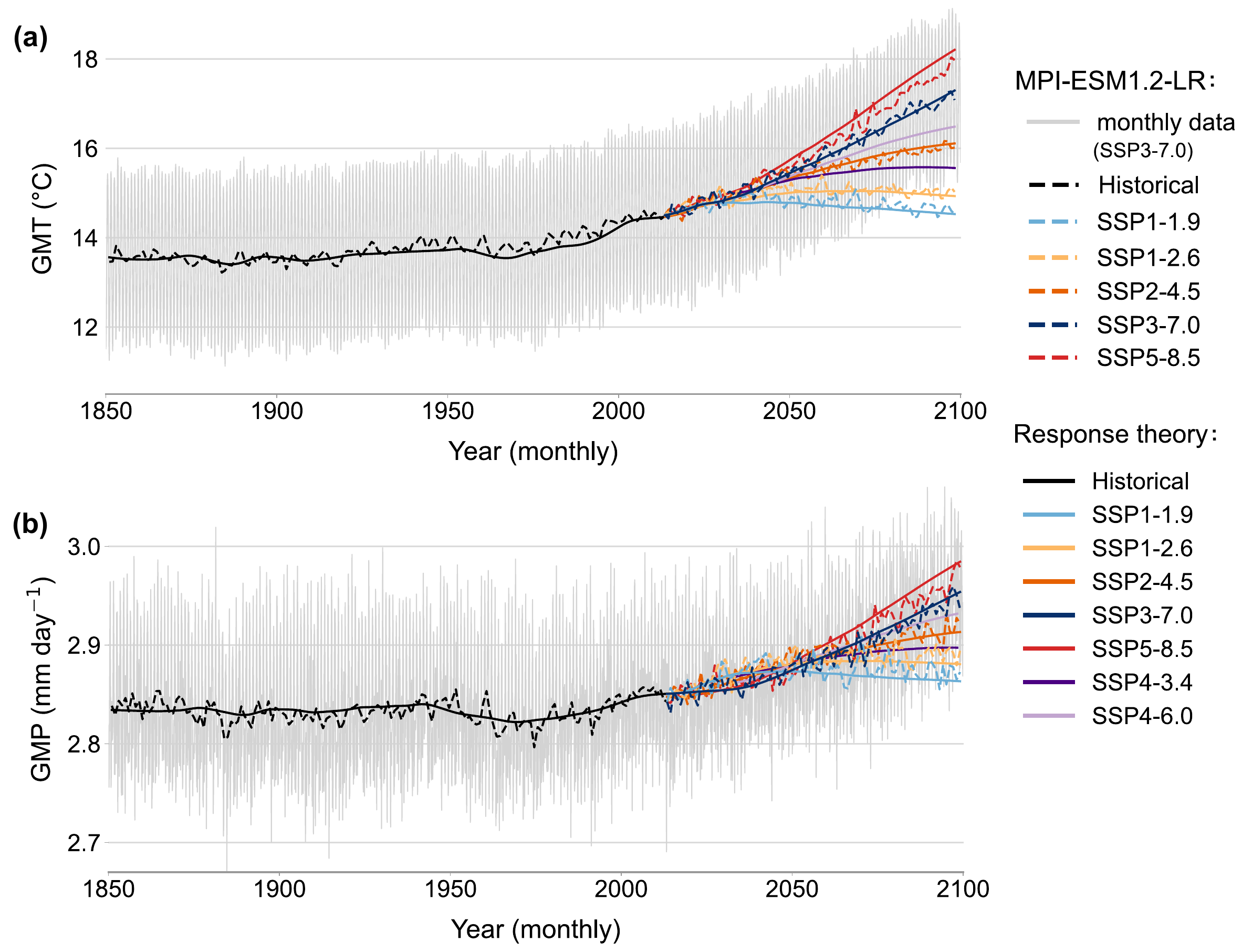}
\caption{\textbf{Global-mean time series of response theory predicted temperature and precipitation under different SSP radiative forcing scenarios.} For MPI-ESM1.2-LR, the response theory model predicts monthly temperature and precipitation fields over the historical period (1850–2014) and across scenarios of SSP1, SSP2,SSP3, SSP5 (2015–2100), with extended projections for SSP4-3.4 and SSP4-6.0. A comparison between native ESM data and response theory predictions is presented through the time series of globally averaged temperature ($\mathbf{a}$) and precipitation ($\mathbf{b}$). To visualize the long-term temporal trends in different SSP scenarios, the time series shown are smoothed with a 12-month running average. }\label{fig2}
\end{figure}

\clearpage
\begin{figure}[htbp]
\centering
\includegraphics[width=1.01\textwidth]{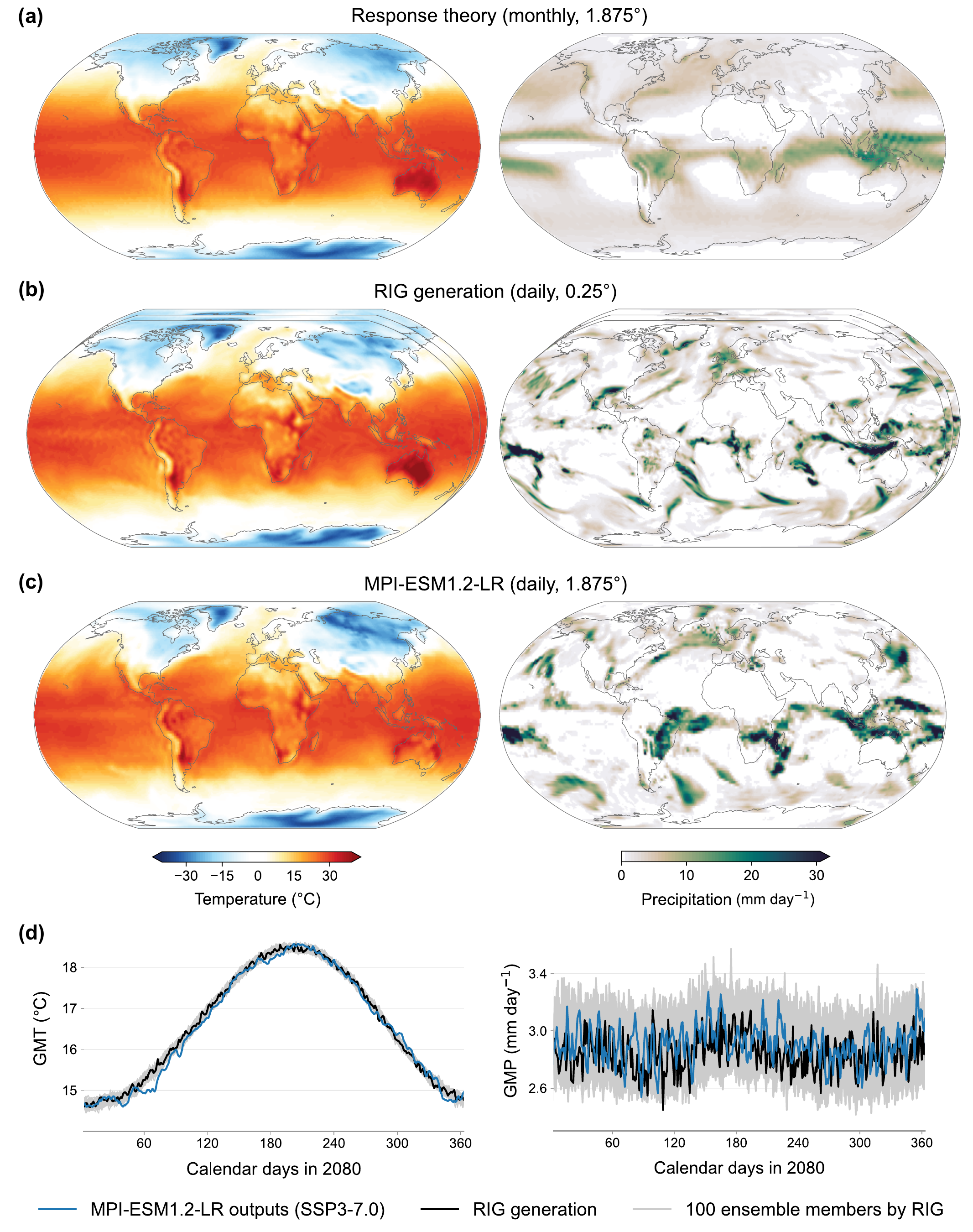}
\caption{\textbf{Generated daily temperature and precipitation fields under the guidance of response theory.} $\mathbf{a\text{-}b}$, Conditioned on the monthly fields predicted by response theory for January 2080 ($\mathbf{a}$), RIG generates consecutive daily fields ($\mathbf{b}$) and downscales the spatial resolution from 1.875° to 0.25°. 
$\mathbf{c}$, Daily fields simulated by MPI-ESM1.2-LR for January 2080 at a spatial resolution of 1.875° for comparison.
$\mathbf{d}$, A comparison between MPI-ESM1.2-LR and RIG is presented through daily time series of globally averaged temperature and precipitation for 2080. Results from 100 ensemble members generated by RIG are shown.
}\label{fig3}
\end{figure}

\clearpage
\begin{figure}[htbp]
\centering
\includegraphics[width=1.05\textwidth]{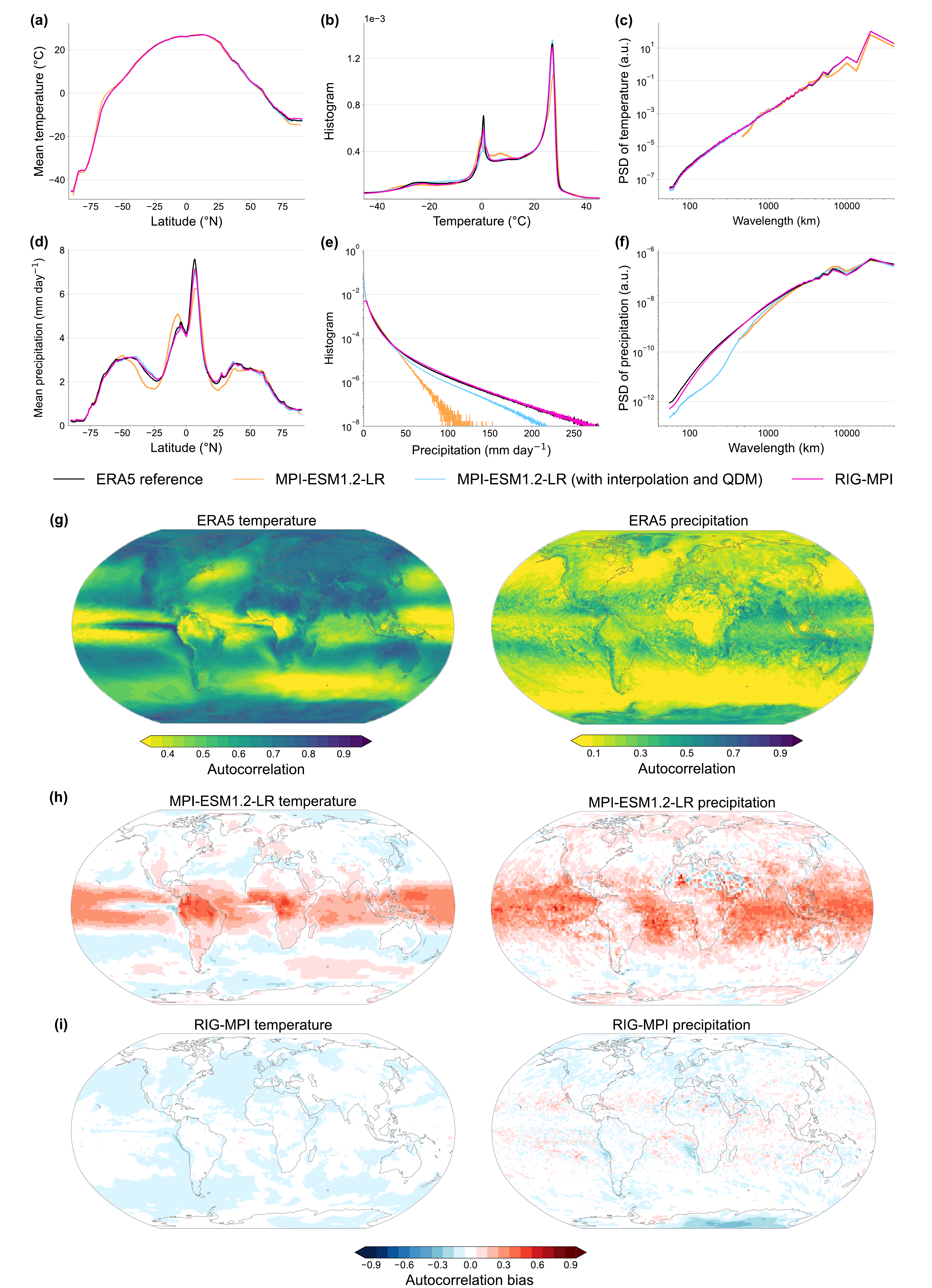}
\caption{\textbf{Quantitative evaluation of the statistical properties of generated daily fields during 2015-2023.} $\mathbf{a\text{-}c}$, The daily temperature fields of RIG-MPI are compared with the ERA5 reference, MPI-ESM1.2-LR, and the post-processed counterpart of MPI-ESM1.2-LR in terms of longitude mean distribution ($\mathbf{a}$), relative histogram ($\mathbf{b}$), and spatial power spectral density (PSD) ($\mathbf{c}$). All results are based on the temporal average from  the period 2015 to 2023 unseen during training. $\mathbf{d\text{-}f}$, Same as $\mathbf{a\text{-}c}$, but for precipitation. $\mathbf{g}$, The lag-1 autocorrelation coefficients of daily temperature (left) and precipitation (right) anomalies from the ERA5 reference for individual grid cells on the global map. $\mathbf{h}$, The difference of lag-1 autocorrelation coefficients between MPI-ESM1.2-LR and the ERA5 reference. $\mathbf{i}$, Same as $\mathbf{h}$, but for the RIG-MPI.  
         }\label{fig4}
\end{figure}

\clearpage
\begin{figure}[htbp]
\centering
\includegraphics[width=1\textwidth]{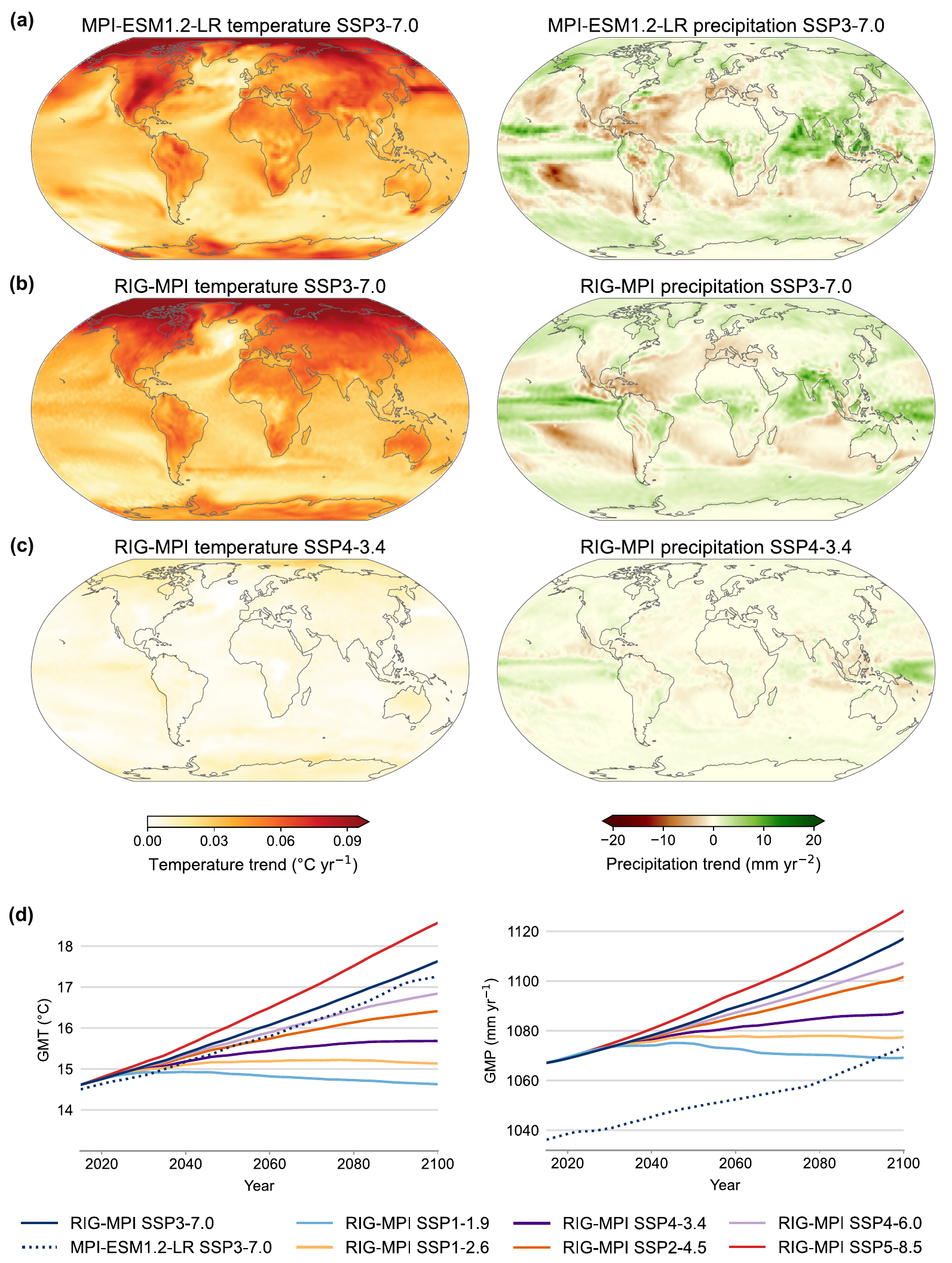}
\caption{\textbf{Long-term trends of temperature and precipitation generated by RIG-MPI in different radiative forcing scenarios.} $\mathbf{a}$, Spatially resolved long-term trends of temperature (left) and precipitation (right) from MPI-ESM1.2-LR under SSP3-7.0. The analysis covers the period from 2050 to 2100. $\mathbf{b}$, Same as $\mathbf{a}$, but for RIG-MPI under SSP3-7.0. $\mathbf{c}$, Same as $\mathbf{a}$, but for RIG-MPI under SSP4-3.4. $\mathbf{d}$, Global mean temperature (left) and precipitation (right) trends from RIG-MPI under seven major SSP scenarios are compared, together with the SSP3-7.0 simulation from MPI-ESM1.2-LR. The time series are smoothed with a 20-year running mean. The systematic shift between MPI-ESM1.2-LR SSP3-7.0 and the corresponding RIG-MPI is a result of the bias correction as part of RIG.   
         }\label{fig5}
\end{figure}

\clearpage
\begin{figure}[htbp]
\centering
\includegraphics[width=1.05\textwidth]{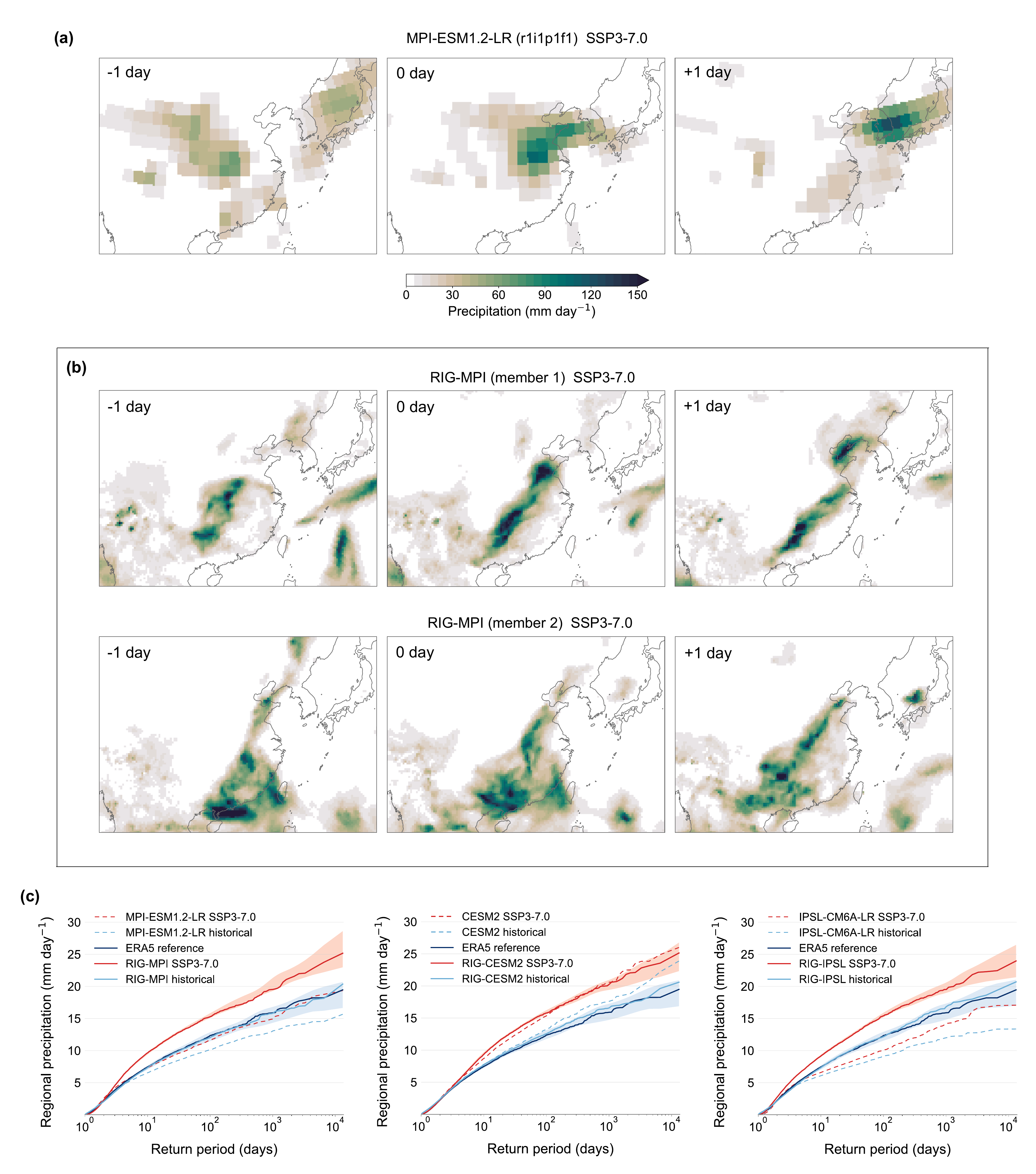}
\caption{\textbf{Visualization of intense precipitation events generated by RIG.} $\mathbf{a}$, The spatial patterns of the most intense extreme precipitation event simulated by MPI-ESM1.2-LR in the Eastern Asia region during 2065-2100 under SSP3-7.0 are presented, with its spatiotemporal evolution over three consecutive days. $\mathbf{b}$, Same as $\mathbf{a}$, but for events from two ensemble members of RIG-MPI with the horizontal resolution enhanced from 1.875° to 0.25°. $\mathbf{c}$, Return period analyses of regionally averaged precipitation from RIG-MPI (left), RIG-CESM2 (middle), and RIG-IPSL (right) are shown. Respective statistics are performed on daily precipitation for the periods 1979-2014 (historical) and 2065-2100 (SSP3-7.0). These are compared with precipitation from the corresponding native ESMs and the ERA5 reference for the same periods. Results from 100 ensemble members generated by RIG are also shown for comparison, with blue and red shaded areas indicating the ensemble spread. }\label{fig6}
\end{figure}

\end{document}